\documentclass[12pt,onecolumn, draftcls]{IEEEtran}

\usepackage{amsmath,amssymb,citesort}
\usepackage{graphicx}
\usepackage{subfigure}
\usepackage{color,setspace}

\newtheorem{theorem}{Theorem}
\newtheorem{proposition}{Proposition}
\newtheorem{lemma}{Lemma}

\newtheorem{remark}{Remark}

\begin{document}

\title{Multiuser Joint Energy-Bandwidth Allocation with Energy Harvesting - Part I: Optimum  Algorithm \& Multiple Point-to-Point Channels}

\author{Zhe~Wang,
        Vaneet~Aggarwal, 
        Xiaodong~Wang
\thanks{Z. Wang and X. Wang are with the Electrical Engineering Department, Columbia University, New York, NY 10027 (e-mail:
\{zhewang, wangx\}@ee.columbia.edu).}
\thanks{V. Aggarwal is with AT\&T Labs-Research, Bedminster, NJ 07921 USA (e-mail: vaneet@research.att.com).}
\thanks{This paper was presented in part at the 2014 IEEE International Symposium on Information Theory (ISIT), Honolulu, HI, USA, Jul. 2014.}
}

\maketitle

\begin{abstract}
In this paper, we develop optimal energy-bandwidth allocation algorithms in fading channels for multiple  energy harvesting transmitters, each may communicate with multiple receivers via orthogonal channels. We first assume that the side information of both the channel states and the energy harvesting states is known for $K$ time slots {\em a priori}, and the battery capacity and the maximum transmission power in each time slot are bounded. The objective is to maximize the weighted sum-rate of all transmitters over the $K$ time slots by assigning the transmission power and bandwidth for each transmitter in each slot. The problem is formulated as a convex optimization problem with ${\cal O}(MK)$ constraints, where $M$ is the number of the receivers, making it hard to solve with a generic convex solver. An iterative algorithm is proposed that alternatively solves two subproblems in each iteration. The convergence and the optimality of this algorithm are also shown. We then consider the special case that each transmitter only communicates with one receiver and the objective is to maximize the total throughput. We develop efficient algorithms for solving the two subproblems and the optimal energy-bandwidth allocation can be obtained with an overall complexity of ${\cal O}(MK^2)$. Moreover, a heuristic algorithm is also proposed for energy-bandwidth allocation based on causal information of channel and energy harvesting states. 

\end{abstract}


\IEEEpeerreviewmaketitle

\section{Introduction}
With the rapid development of energy harvesting technologies, a new paradigm of wireless communications that employs energy harvesting transmitters has become a reality \cite{EHSNSI}. The renewable energy source enables the flexible deployment of the transmitters and prolongs their lifetimes \cite{EHSNSI}\cite{Energy_Scav}. State-of-the-art techniques can provide fairly accurate short-term prediction of the energy harvesting process, which can be used to assist energy scheduling \cite{WPDFUE}\cite{PMEHW}. To make the best use of the harvested energy for wireless communications, many challenging research issues arise \cite{ALLERTON}\cite{2014arXiv1401.2376W}\cite{OTPBLE}\cite{TMGRCE}\cite{OPSMAC}. In particular, optimal resource (energy, bandwidth, etc.) scheduling is key to the design of an efficient wireless system powered by renewable energy sources.

For a single transmitter with energy harvesting, a number of works addressed the energy scheduling problem with non-causal channel state information. For static channels, \cite{FHEARS} proposed a shortest-path-based algorithm for the energy scheduling. \cite{OTPBLE} analyzed the optimality properties based on the energy causality and the optimal energy scheduling algorithm was also provided.  For fading channels, a staircase water-filling algorithm was proposed in \cite{OEAWCE} for the case of infinite battery capacity; with finite battery capacity, \cite{TEHNFW} studied the energy flow behavior with an energy harvesting device and proposed a directional water-filling method. Taking  the maximum transmission power into account, \cite{ALLERTON} proposed a dynamic water-filling algorithm to efficiently obtain the energy schedule to maximize the achievable rate. Energy scheduling for multiuser systems with energy harvesting transmitters has also been considered. In \cite{OPSMAC},  the general capacity region for a static multiple-access channel (MAC) was characterized  without considering the constraints on the battery capacity and the maximum transmission power. \cite{SROPPE} discussed the optimal power policy for energy harvesting transmitters in a two-user Gaussian interference channel. In \cite{broadcasting},  the optimal energy scheduling algorithm was proposed for a static broadcast channel  with finite battery capacity constraint.  Considering both the finite  battery capacity and the finite maximum transmission power, the iterative dynamic water-filling algorithm was extended to  the fading MAC channel \cite{2014arXiv1401.2376W}. Moreover, the scheduling problem in the Gaussian relay channel with energy harvesting was discussed in \cite{TMGRCE}.


In this paper, we consider a multiuser system  with multiple transmitters, each powered by a renewable energy source.  Each transmitter communicates with its designated receivers and  is constrained by the availability of the energy, the capacity of the battery, and the maximum (average) transmission power. Moreover, a frequency band is shared by all transmitters and we assume orthogonal channel access to avoid interference. We aim to obtain the optimal joint energy-bandwidth allocation over a fixed scheduling period based on the available information on the channel states and energy harvesting states at all transmitters, to maximize the weighted sum of the achievable rate.

Consider the special case of equal weights and each transmitter communicates with only one receiver.  Then, without energy harvesting, TDMA is optimal for the maximum unweighted sum-rate, i.e., at any time the link with the highest rate takes all bandwidth. However, for energy harvesting transmitters, TDMA is no longer optimal. This is because the finite battery capacity leads to energy discharge and waste by some transmitters that are not scheduled to transmit in a time slot. Therefore, to make the best use of the harvested energy, multiple transmitters should split the frequency band and transmit in a same slot. In this paper, we assume that the channel is flat fading and therefore each transmitter only needs to be allocated a portion of the total bandwidth.

We first consider the non-causal case, i.e., the energy harvesting and the channel fading can be predicted for the scheduling period, and formulate a convex optimization problem with ${\cal O}(MK)$ variables and constraints, where $M$ is the number of receivers and $K$ is the number of scheduling time slots. Since the computational complexity of a generic convex solver becomes impractically high when the number of constraints is large \cite{CO}, we will develop an iterative algorithm that alternates between  energy allocation and bandwidth allocation. We will show that this algorithm converges to the optimal solution of the joint energy-bandwidth scheduling problem. For the special case that each transmitter only communicates with one receiver and all weights are equal, optimal algorithms to solve the energy and bandwidth allocation subproblems are also the optimal energy-bandwidth allocation algorithm that obtained a computational complexity of ${\cal O}(MK^2)$. We then consider the causal case, where the harvested energy and the channel gain can only be observed at the beginning of the corresponding time slot. We propose a suboptimal energy-bandwidth allocation algorithm that follows a similar structure of the noncausal optimal solution. Simulation results demonstrate that both the proposed non-causal and causal algorithms achieve substantial performance improvement over some heuristic scheduling policies.

The remainder of the paper is organized as follows. In Section II, we describe the system model
and formulate the joint  energy-bandwidth scheduling problem. In Section III, we propose an iterative algorithm to obtain the optimal energy-bandwidth allocation and prove its convergence and optimality. Optimal algorithms for energy allocation and bandwidth allocation, respectively,  are further developed in Section IV for the special case that each transmitter only communicates with one receiver and all weights are equal. In Section V, a suboptimal causal algorithm is proposed.  Simulation results are provided in Section VI. Finally, Section VII concludes the paper.

\section{System Model and Problem Formulation}
\subsection{System Model}
Consider a network consisting of $N$ transmitters and $M$ receivers sharing  a total   bandwidth of $B$ Hz, where $N\leq M$ and each transmitter may communicate with multiple receivers.  We assume a scheduling period of $K$ time slots and no two transmitters can transmit in the same time slot and the same frequency band. Denote $a_{m}^k\in[0,1]$ as the normalized bandwidth allocation for link $m$ in time slot $k$. We consider a flat and slow fading channel, where the channel gain is constant within the entire frequency band of $B$ Hz  and over the coherence time of $T_c$ seconds, which is also the duration of a time slot. Assuming that each time slot consists of $T$ time instants, we denote $X_{mki}$ as the symbol sent to the receiver of link $m$ at instant $i$ in slot $k$. The corresponding  received signal at receiver $m$  is given by
\begin{equation}
Y_{mki} = h_{mk}X_{mki} + Z_{mki}
\end{equation}
where $h_{mk}$ represents the complex channel gain for link $m$ in slot $k$,  and $Z_{mki}\sim {\sf CN}(0,1)$ is the i.i.d. complex Gaussian noise. We denote $H_m^k\triangleq |h_{mk}|^2$ and denote $p_m^k\triangleq \frac{1}{T_c}\sum_{i}|X_{mki}|^2$ as the transmission energy consumption for link $m$ in slot $k$. Without loss of generality, we normalize both $T_c$ and $B$ to $1$; then, $p_m^k$ and $a_m^k$ become the transmission power and the allocated bandwidth of link $m$ in slot $k$, respectively. For link $m$, the upper bound of the achievable channel rate in slot $k$ can be written as $a_m^k\log(1+p_m^kH_m^k/a_m^k)$ \cite{IT}. Moreover, we denote ${\cal K}\triangleq \{1,2,\ldots,K\}$ as the scheduling period, ${\cal N}\triangleq \{1,2,\ldots,N\}$ as the set of transmitters, and ${\cal M}\triangleq \{1,2,\ldots,M\}$ as the set of receivers. Further, we denote ${\cal M}_n\triangleq\{m\;|\; m \textrm{ is the receiver of transmitter } n, m\in{\cal M}\}$ as the set of receivers of transmitter $n$, where ${\cal M}_n \bigcap {\cal M}_{n'}=\phi$ for all $n\neq n'\in{\cal N}$.

\begin{figure}[!hbt]
\centering
\includegraphics[width=.65\textwidth]{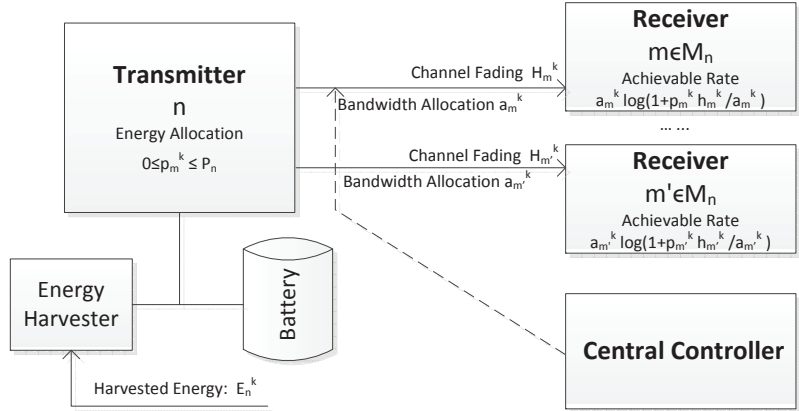}
\caption{The system block diagram.}
\label{fg:system}
\end{figure}

Assume that each transmitter is equipped with an energy harvester and a buffer battery, as shown in Fig.~\ref{fg:system}. The energy harvester harvests energy from the surrounding environment. We denote $E_n^k$ as the total energy harvested up to the end of slot $k$ by transmitter $n$. Since in practice energy harvesting can be accurately predicted for a short period~\cite{WPDFUE}\cite{PMEHW}, we assume that the amount of the harvested energy in each slot is known.  Moreover, the short-term prediction of the channel gain in slow fading channels is also possible \cite{FCPMRA}. Therefore, we assume that $\{H_m^k\}$ and $\{E_n^k\}$ are known non-causally before scheduling. Note that such non-causal assumption also leads to the performance upper bound of the system.  We will relax this assumption and consider causal knowledge of the channels and energy harvesting  in Section V.

For transmitter $n$, assuming that the battery has a limited capacity $B_n^{\max}$ and is empty initially, then the battery level at the end of slot $k$ can be written as
\begin{equation}\label{eq:battery}
B_n^k = B_n^{k-1} + \left( E_n^k - E_n^{k-1}\right) -\sum_{\kappa=1}^{k}\sum_{m\in{\cal M}_n}p_m^{\kappa}-\sum_{\kappa=1}^{k}D_n^{\kappa},
\end{equation}
where $D_n^k\geq 0$ represents the energy discharge (waste) for transmitter $n$ in slot $k$. Moreover, $B_n^k$ must satisfy  $0\leq B_n^k \leq B_n^{\max}$ for all $k\in{\cal K}$.

We assume that each transmitter $n$ has a maximum per-slot transmission energy consumption, $P_n$, such that $\sum_{m\in{\cal M}_n}p_m^k\leq P_n$ for $k\in{\cal K}$. With the maximum transmission energy and the limited battery capacity, some of the harvested energy may not be able to be utilized, and is therefore wasted, i.e., $D_n^k$ may necessarily be strictly positive in some slots. Then, the constraints on the battery level can be written as
\begin{equation}\label{eq:rbattery}
0\leq  E_n^k - \sum_{\kappa=1}^{k}\sum_{m\in{\cal M}_n}p_m^{\kappa} -\sum_{\kappa=1}^{k}D_n^{\kappa} \leq B_n^{\max}.
\end{equation}
Moreover, we denote ${\cal D}\triangleq \{\boldsymbol{D}_n\;|\; \boldsymbol{D}_n\triangleq [D_n^1,D_n^2,\ldots,D_n^K],n\in{\cal N}\}$ as the {\em discharge allocation}. Note that, we assume controllable energy discharge, i.e., the energy can be discharged and wasted anytime, even when the battery is not full.

\begin{remark}
In the transmitter model, both the maximum transmission energy and the battery capacity are finite. If the harvested energy is ample, part of the energy has to be discharged even if the transmitter transmits at the maximum (available) transmission energy in each slot. That is, ${D_n^k}^*>0$ is due to the incoming energy being large enough that it cannot be used for transmission or storage.
\end{remark}

\subsection{Problem Formulation}
Define $0\cdot\log(1+\frac{x}{0}) \triangleq 0$. We use upper bounds of the achievable channel rate over a weighted sum of the $M$ links and $K$ slots as the performance metric, given by
\begin{equation}\label{eq:oobj}
C_{\cal W}({\cal P},{\cal A}) = \sum_{m\in{\cal M}}W_{m}\sum_{k\in{\cal K}} a_m^k\cdot \log(1+\frac{p_m^kH_m^k}{a_m^k}),\ a_m^k\in[0,1], p_m^k\in[0,\infty)\ ,
\end{equation}
where ${\cal P}\triangleq \{\boldsymbol{p}_m\;|\; \boldsymbol{p}_m\triangleq [p_m^1,p_m^2,\ldots,p_m^K],m\in{\cal M}\}$ is the {\em energy allocation},   ${\cal A}\triangleq \{\boldsymbol{a}^k\;|\; \boldsymbol{a}^k\triangleq [a_1^k,a_2^k,\ldots,a_M^k],k\in{\cal K}\}$ is the {\em bandwidth allocation}, and ${\cal W}\triangleq \{W_m,m\in{\cal M}\}$ is the {\em weight set}. In particular, when $W_m=1$ for all $m\in{\cal M}$, $C_{\cal W}({\cal P},{\cal A}) $ becomes the throughput of the network.

Note that, both $a_m^k$ and $p_m^k$ can be zero in \eqref{eq:oobj}. However, if $a_m^k=0$, the channel rate of link $m$ in slot $k$ is zero, even if the energy allocation $p_m^k>0$, thus $p_m^k$ is actually wasted. However, we still treat the pair $(a_m^k=0,p_m^k >0)$ as feasible as long as $\sum_{m\in{\cal M}_n}p_m^k \leq P_n$.


We formulate the following energy-bandwidth allocation problem:
\begin{equation}\label{eq:oproblem}
\max_{{\cal P},{\cal A},{\cal D}}C_{\cal W}({\cal P},{\cal A})
\end{equation}
subject to
\begin{equation}\label{eq:ocst}
\left\{\begin{array}{l}
0\leq E_n^k - \sum_{\kappa=1}^{k}\sum_{m\in{\cal M}_n}p_m^{\kappa}-\sum_{\kappa=1}^{k}D_n^{\kappa}\leq B_n^{\max}\\
\sum_{i=1}^M a_i^k = 1\\
a^k_m \geq 0\\
\sum_{m\in{\cal M}_n} p_m^k \leq P_n\\
p_m^k \geq 0\\
D_n^k\geq 0
\end{array}\right.
\end{equation}
for all $k\in{\cal K}, m\in{\cal M}$ and $n\in{\cal N}$.

\subsection{Optimal Energy Discharge Allocation}

To efficiently solve the problem in \eqref{eq:oproblem}-\eqref{eq:ocst}, we consider a two-stage procedure. In the first stage, we obtain the optimal energy discharge allocation ${\cal D}^*$ such that
\begin{equation}\label{eq:optdischarg}
\max_{{\cal P},{\cal A},{\cal D}={\cal D}^*}C_{\cal W}({\cal P},{\cal A}) =\max_{{\cal P},{\cal A},{\cal D}}C_{\cal W}({\cal P},{\cal A})
\end{equation}
with the constraints in \eqref{eq:ocst}. And in the second stage, we use ${\cal D}^*$ and define the energy expenditure for transmission as
\begin{equation}
\tilde{E}_n^k\triangleq {E}_n^k-\sum_{\kappa=1}^{k}{{D}_n^{\kappa}}^*\ .
\end{equation}
Then we solve the following problem:
\begin{equation}\label{eq:problem}
\max_{{\cal P},{\cal A}}C_{\cal W}({\cal P},{\cal A})
\end{equation}
subject to
\begin{equation}\label{eq:cst}
\left\{
\begin{array}{l}
\tilde{E}_n^k - B_n^{\max} \leq  \sum_{\kappa=1}^{k}\sum_{m\in{\cal M}_n}p_m^{\kappa} \leq \tilde{E}_n^k \\
\sum_{i=1}^M a_i^k = 1\\
\sum_{m\in{\cal M}_n} p_m^k \leq P_n\\
p_m^k \geq 0\\
a_m^k \geq 0\\
\end{array}\right.
\end{equation}
for all $n\in{\cal N}, m\in{\cal M}$ and $k\in{\cal K}$.

We consider the following greedy strategy to obtain the energy discharge allocation by assuming that each transmitter transmits at the maximum power in each slot, i.e.,
\begin{equation}\label{eq:greedy}
\left\{\begin{array}{l}
D_n^k = \max\{B_n^{k-1} + E_n^k-E_n^{k-1}- \sum_{\kappa=1}^{k}\sum_{m\in{\cal M}_n}p_m^k-B_n^{\max},0\}\\
\sum_{m\in{\cal M}_n}p_m^k = \min\{P_n,B_n^{k-1}+E_n^k-E_n^{k-1}\}
\end{array}\right.,\ k=1,2,\ldots, K
\end{equation}
for all $n\in{\cal N}$.

Note that, following \eqref{eq:greedy}, the total discharged energy is minimized and thus the amount of the energy used for transmission is maximized. Intuitively, this way the feasible domain becomes the largest, providing the best performance for transmission energy scheduling. Specifically, given a feasible bandwidth allocation $\cal A$, the achievable rate of each link is non-decreasing with respect to the transmission energy, and the battery of each transmitters operates independently. Therefore,  following the same lines of the proof in \cite{2014arXiv1401.2376W}, the optimality of \eqref{eq:greedy} can be established. In particular, using any feasible energy discharge $\cal D$ corresponding to the minimal energy wastage, the optimal value of \eqref{eq:problem} is same, which is no less than the optimal value under any feasible energy discharge allocation with non-minimal energy wastage.

\begin{lemma}
The discharge allocation given by \eqref{eq:greedy} is the optimal ${\cal D}^*$ to the problem in \eqref{eq:oproblem}-\eqref{eq:ocst}, i.e., it satisfies \eqref{eq:optdischarg}, where the LHS of \eqref{eq:optdischarg} is subject to the constraints in \eqref{eq:cst} and the RHS is subject to the constraints in \eqref{eq:ocst}.
\end{lemma}

Note that, $C_{\cal W}({\cal P},{\cal A})$ is continuous and jointly concave with respect to  $a_m^k\in[0,1]$ and  $p_m^k\in[0,\infty)$ for $k\in{\cal K},m\in{\cal M}$. Then, the problem in \eqref{eq:problem}-\eqref{eq:cst} is a convex optimization problem and can be solved by a generic convex solver, whose complexity becomes impractically high when the number of constraints is large \cite{CO}, which in this case is ${\cal O}(MK)$. To reduce the computational complexity, we will develop an efficient algorithm in this paper, which exploits the structure of the optimal solution.



\subsection{K.K.T. Conditions for Non-Zero Bandwidth Allocation}
The problem in \eqref{eq:problem}-\eqref{eq:cst} is a convex optimization problem with linear constraints. When the objective function is differentiable in an open domain, the K.K.T. conditions are sufficient and necessary for the optimal solution \cite{CO}. Note that, \eqref{eq:oobj} is non-differentiable at $a_m^k=0$. To use the K.K.T. conditions to characterize the optimality of the problem in \eqref{eq:problem}-\eqref{eq:cst}, we consider the following  approximation:
\begin{equation}\label{eq:aproblem}
{\sf P}_{\cal W}(\epsilon):\quad \max_{{\cal P},{\cal A}}C_{\cal W}({\cal P},{\cal A})
\end{equation}
subject to
\begin{equation}\label{eq:acst}
\left\{
\begin{array}{l}
\tilde{E}_n^k - B_n^{\max} \leq  \sum_{\kappa=1}^{k}\sum_{m\in{\cal M}_n}p_m^{\kappa} \leq \tilde{E}_n^k \\
\sum_{i=1}^M a_i^k = 1\\
\sum_{m\in{\cal M}_n} p_m^k \leq P_n\\
p_m^k \geq 0\\
a_m^k \geq \epsilon
\end{array}\right.
\end{equation}
for all $n\in{\cal N}, m\in{\cal M}, k\in{\cal K}$, where $\epsilon$ is a small positive number. In particular, ${\sf P}_{\cal W}(0)$ is the original problem in \eqref{eq:problem}-\eqref{eq:cst}.

\begin{lemma}\label{lm:approx}
When $\epsilon\rightarrow 0^+$, the optimal value of ${\sf P}_{\cal W}(\epsilon)$ converges to the optimal value of the problem in \eqref{eq:problem}-\eqref{eq:cst}, i.e., $\lim_{\epsilon\rightarrow 0^+}{\sf P}_{\cal W}(\epsilon) = {\sf P}_{\cal W}(0)$.
\end{lemma}
\begin{IEEEproof}
Since the objective function $C_{\cal W}({\cal P,A})$ is continuous with respect to ${\cal P}\times{\cal A}\in\{[0,\infty)\}\times\{[0,1]\}$ and the constraints in \eqref{eq:acst} are all linear, we have that the optimal solution of ${\sf P}_{\cal W}(\epsilon)$ is continuous with respect to $\epsilon$, i.e., $\lim_{\epsilon\rightarrow 0^+}\arg{\sf P}_{\cal W}(\epsilon) =\arg {\sf P}_{\cal W}(0)$. Therefore, we have $\lim_{\epsilon\rightarrow 0^+}{\sf P}_{\cal W}(\epsilon) = {\sf P}_{\cal W}(0)$.
\end{IEEEproof}

By introducing the auxiliary variables $\{\lambda_n^k \geq 0\}$, $\{\mu_n^k\geq 0\}$, $\{\beta_m^k \geq 0\}$ and $\{\alpha^k\}$ and converting the constraints in \eqref{eq:acst} into the Lagrangian multiplier, we can define the Lagrangian function for ${\sf P}_{\cal W}(\epsilon)$ as
\begin{align}
{\cal L}\triangleq&\;\; \sum_{m=1}^MW_{m}\sum_{k=1}^Ka_m^k\cdot\log(1+\frac{p_m^kH_m^k}{a_m^k}) \nonumber \\
& -\sum_{n=1}^N\sum_{k=1}^K \Big( \sum_{m\in{\cal M}_n}p_m^k \sum_{\kappa=k}^K\lambda_n^{\kappa}- \lambda_n^k\tilde{E}_n^k\Big) \nonumber \\
& +\sum_{n=1}^N\sum_{k=1}^K \Big(  \sum_{m\in{\cal M}_n}p_m^k\sum_{\kappa=k}^K\mu_n^{\kappa}- \mu_n^k \big(\tilde{E}_n^k-B_n^{\max}\big)\Big) \nonumber \\& -  \sum_{k=1}^K\alpha^k (\sum_{m=1}^M a_m^k - 1)  \nonumber \\&+  \sum_{k=1}^K\sum_{m=1}^M \beta_m^k (a_m^k - \epsilon) \nonumber\ .
\end{align}

Then, the following K.K.T. conditions, which are sufficient and necessary for the optimal solution to the convex optimization problem in \eqref{eq:aproblem}-\eqref{eq:acst},  are obtained from the Lagrangian function:
\begin{align}
\frac{H_m^k}{1+p_m^kH_m^k/a_m^k} = (v_n^k - u_n^k)/W_m,&\ k\in{\cal K},\ n\in{\cal N},\ m\in{\cal M}_n\label{kkt:wf}\\
\log(1+\frac{p_m^kH_m^k}{a_m^k}) - \frac{p_m^kH_m^k}{a_m^k + p_m^kH_m^k} = (\alpha^k-\beta_n^k)/W_m,&\ k\in{\cal K},\ n\in{\cal N},\ m\in{\cal M}_n\label{kkt:al}\\
\lambda_n^k\cdot(\sum_{\kappa=1}^k\sum_{m\in{\cal M}_n}p_m^{\kappa}-{\tilde E}_n^k)=0,&\ k\in{\cal K},\ n\in{\cal N}\label{kkt:a}\\
\mu_n^k\cdot (\sum_{\kappa=1}^k\sum_{m\in{\cal M}_n}p_m^{\kappa}-{\tilde E}_n^k+B_n^{\max})=0,&\ k\in{\cal K},\ n\in{\cal N}\label{kkt:b}\\
\alpha^k\cdot (\sum_{m=1}^M a_m^k - 1) = 0,&\ k\in{\cal K}\label{kkt:c}\\
\beta_m^k\cdot ( a_m^k - \epsilon) = 0,&\ k\in{\cal K},\ m\in{\cal M}\label{kkt:d}
\end{align}
together with the constraints in \eqref{eq:acst}, and $\lambda_n^k,\mu_n^k,\beta_m^k\geq 0$ for all $k\in{\cal K}$, $n\in{\cal N}$, and $m\in{\cal M}$, where in \eqref{kkt:wf}
\begin{align}
u_n^k \triangleq \sum_{\kappa=k}^K\mu_n^{\kappa},\ v_n^k \triangleq \sum_{\kappa=k}^K\lambda_n^{\kappa}\label{eq:l2}\ .
\end{align}

\section{Iterative Algorithm and its Optimality}
In this section, we will first decompose the energy-bandwidth allocation problem ${\sf P}_{\cal W}(\epsilon)$ in \eqref{eq:aproblem}-\eqref{eq:acst} into two subproblems, and then propose an iterative algorithm to solve ${\sf P}_{\cal W}(\epsilon)$. We will prove that the iterative algorithm converges to the optimal solution to the problem in \eqref{eq:problem}-\eqref{eq:cst}.

\subsection{Iterative Algorithm}
To efficiently solve  problem ${\sf P}_{\cal W}(\epsilon)$ in \eqref{eq:aproblem}-\eqref{eq:acst}, we first decompose it into two groups of subproblems, corresponding to energy allocation and bandwidth allocation, respectively.

\begin{itemize}
\item Given the bandwidth allocation ${\cal A} = \{ \boldsymbol{a}^k\; |\; k \in {\cal K} \}$, for each $n \in{\cal N}$, obtain the energy allocation $\boldsymbol{p}_m$ by solving the following subproblem:
\begin{equation}
{\sf EP}_n:\quad \max_{\boldsymbol{p}_m,m\in{\cal M}_n} \sum_{m\in{\cal M}_n}W_m \sum_{k=1}^K  a_m^k\cdot \log(1+\frac{p_m^kH_m^k}{a_m^k})
\end{equation}
subject to
\begin{equation}
\left\{
\begin{array}{l}
\tilde{E}_n^k - B_n^{\max} \leq  \sum_{\kappa=1}^{k}\sum_{m\in{\cal M}_n}p_m^{\kappa} \leq \tilde{E}_n^k \\
\sum_{m\in{\cal M}_n} p_m^k \leq P_n\\
p_m^k \geq 0, m\in{\cal M}_n
\end{array}\right. ,\ k\in{\cal K}\ .
\end{equation}
\item Given the energy allocation ${\cal P} = \{ \boldsymbol{p}_m\; |\; m \in {\cal M} \}$, for each $k \in{\cal K}$, obtain the bandwidth allocation $\boldsymbol{a}^k$ by solving the following subproblem:
\begin{equation}
{\sf BP}_k(\epsilon):\quad\max_{\boldsymbol{a}^k}\sum_{m=1}^M W_m\cdot a_m^k\cdot \log(1+\frac{p_m^kH_m^k}{a_m^k})
\end{equation}
subject to
\begin{equation}
\left\{
\begin{array}{l}
\sum_{i=1}^M a_i^k = 1\\
a_m^k \geq \epsilon,\ m\in{\cal M}
\end{array}\right. \ .
\end{equation}
\end{itemize}


\begin{figure}
\centering
\includegraphics[width=0.45\textwidth]{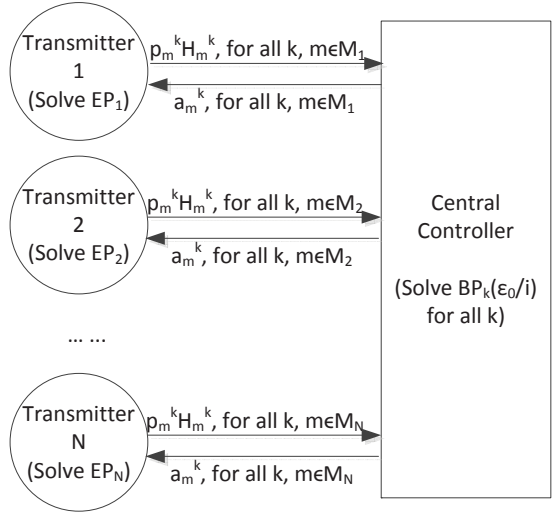}
\caption{The block diagram of Algorithm 1.}
\label{fg:alg}
\end{figure}

To obtain the optimal solution to the original problem in \eqref{eq:problem}-\eqref{eq:cst}, we propose an iterative algorithm that alternatively solves {\sf EP}$_n$ for all $n\in{\cal N}$ and {\sf BP}$_k(\epsilon)$ for all $k\in{\cal K}$,
with a diminishing $\epsilon$ over the iterations. To perform the algorithm, we initially set $a_m^k=1/M,\forall m,k$, and solve ${\sf EP}_n$ to obtain the initial $\cal P$. In each iteration $i$, we first solve {\sf BP$_k(\epsilon_0/i)$} to update $\boldsymbol{a}^k\in{\cal A}$ for all  $k\in{\cal K}$, where $\epsilon_0$ is a pre-specified positive value; with the updated ${\cal A}$, we then solve {\sf EP$_n$} to update  $\boldsymbol{p}_m\in{\cal P}$ for all $m\in{\cal M}$. 

The proposed iterative algorithm is summarized in Algorithm 1 and its block diagram is shown in Fig.~\ref{fg:alg}.\\
\begin{minipage}[h]{6.5 in}
\rule{\linewidth}{0.3mm}\vspace{-.1in}
{\bf {\footnotesize Algorithm 1 - Iterative Energy-Bandwidth Allocation Algorithm}}\vspace{-.2in}\\
\rule{\linewidth}{0.2mm}
{ {\small
\begin{tabular}{ll}
	\;1:&  Initialization\\
	\;& $i=0$, ${\cal A}=1/M$, $V^{(0)}=0$, Choose any $\epsilon_0>0$, Solve {\sf EP$_n$} for all $n\in{\cal N}$ to generate the initial  ${\cal P}$\\
	\;& Specify the maximum number of iterations $I$, the convergence tolerance $\delta>0$\\
    \;2:& Energy-Bandwidth Allocation\\
    \;& {\bf REPEAT}\\
      \;& \quad $i\leftarrow i+1$, $\epsilon \leftarrow \epsilon_0/i$\\
      	\;& \quad Solve {\sf BP$_k(\epsilon)$} to update $\boldsymbol{a}^k\in{\cal A}$ for all  $k\in{\cal K}$\\
      	\;& \quad Solve {\sf EP$_n$} to update $\{\boldsymbol{p}_m\;|\;m\in{\cal M}_n\}\subset{\cal P}$ for all $n\in{\cal N}$\\
    	\;& \quad $V^{(i)}=C_{\cal W}({\cal P},{\cal A})$\\
	\;& {\bf UNTIL} $|V^{(i)}-V^{(i-1)}|  < \delta$ {\bf OR} $i= I$\\
\end{tabular}}}\\
\rule{\linewidth}{0.3mm}
\end{minipage}\vspace{.01 in}\\

In the next subsection, we will show that Algorithm 1 converges and the pairwise optimal $\cal A$ and $\cal P$ can be obtained, which is also the optimal solution to the problem in \eqref{eq:problem}-\eqref{eq:cst}.

We note that, ${\cal P}_{\cal W}(\epsilon)$ is a convex optimization problem with $O(MK)$ variables and constraints. The computational complexity of using the generic convex solver is non-linear with respect to the number of the variables and constraints, which may be impractically high when $M$ and $K$ become large. Using Algorithm 1, the optimal solution to ${\cal P}_{\cal W}(\epsilon)$ can be obtained by solving ${\cal O}(N+K)$ convex optimization subproblems which contains ${\cal O}(K|{\cal M}_n|)$ or ${\cal O}(M)$ variables and constraints. Therefore, the overall computational complexity can be significantly reduced with Algorithm 1 for large $M$ and $K$.

\subsection{Proof of Optimality}
We first give the following proposition.
\begin{proposition}\label{pp:unique}
Given any bandwidth allocation $\{a_m^k>0\;|\;k\in{\cal K}\},m\in{\cal M}$, the optimal energy allocation for the problem {\sf EP}$_n$ is unique. Also, given the energy allocation $\{p_m^k\;|\;m\in{\cal M}\}, k\in{\cal K}$ such that $\sum_{m=1}^M p_m^k > 0$, the corresponding optimal bandwidth allocation for the problem {\sf BP}$_k(\epsilon)$ is unique.
\end{proposition}
\begin{IEEEproof}
This proposition can be obtained by verifying the strict concavity of $C_{\cal W}({\cal P},{\cal A})$ with respect to ${\cal P}$ given ${\cal A}$, and with respect to ${\cal A}$ given ${\cal P}$.
\end{IEEEproof}

Given a pair $({\cal P},{\cal A})$, if $\boldsymbol{p}_m\in{\cal P}$ is the optimal solution to {\sf EP}$_n$ for all $n\in{\cal N}$ given ${\cal A}$, and $\boldsymbol{a}^k\in{\cal A}$ is the optimal solution to {\sf BP}$_k(\epsilon)$ for all $k\in{\cal K}$ given ${\cal P}$, we say that ${\cal P}$ and ${\cal A}$ are pairwise optimal for ${\sf P}_{\cal W}(\epsilon)$. We also note that, for each subproblem, its K.K.T. conditions form a subset of those of ${\sf P}_{\cal W}(\epsilon)$ given the other primal variables, where any two subsets contain no common dual variable. Then, if the primal variables are pairwise optimal, the K.K.T. conditions in each corresponding subset are satisfied and hence all K.K.T. conditions of ${\sf P}_{\cal W}(\epsilon)$ are satisfied, i.e., the pairwise optimal solution is also the optimal energy-bandwidth allocation for ${\sf P}_{\cal W}(\epsilon)$.
\begin{theorem}\label{thm:jointopt}
The energy-bandwidth allocation $\{{\cal P},{\cal A}\}$ is the optimal solution to ${\sf P}_{\cal W}(\epsilon)$ for any $\epsilon>0$, if and only if, $ \{\boldsymbol{p}_m,m\in{\cal M}_n\}\in{\cal P}$ is optimal to ${\sf P}_{\cal W}(\epsilon)$ given $\{{\cal P}\backslash \{\boldsymbol{p}_m,m\in{\cal M}_n\},{\cal A}\}$ for all $n\in{\cal N}$, and $\boldsymbol{a}^k\in{\cal A}$ is optimal to ${\sf P}_{\cal W}(\epsilon)$ given $\{{\cal A}\backslash \boldsymbol{a}^k,{\cal P}\}$ for all $k\in{\cal K}$.
\end{theorem}


We note that, for ${\sf BP}_k(\epsilon)$, when $\sum_{m=1}^M p_m^k = 0$, the objective value is zero for all feasible bandwidth allocations. Therefore, we can fix $a_m^k = 1/M$ as the optimal bandwidth allocation for this case in Algorithm 1. Then, by Proposition \ref{pp:unique}, we have that the optimal solution to each subproblem in Algorithm 1 is unique. The next theorem establishes the optimality of Algorithm 1. The proof is given in Appendix A.

\begin{theorem}\label{pp:cvg}
Algorithm 1 converges; and the converged solution $(\cal P,\cal A)$ is the optimal solution to the problem in \eqref{eq:problem}-\eqref{eq:cst}.
\end{theorem}

Note that, the convergence is due to the expansion of the feasible domain by reducing $\epsilon$ resulting in the increasing objective value over iterations. The optimality can be proved by first verifying the pairwise optimality of the solution upon convergence and then showing it cannot be suboptimal.


\section{Special Case: Throughput Maximization for Multiple Point-to-Point Channels}

In this section, we consider the special case that each transmitter can only communicate with one receiver and all links have the same weight, i.e., ${\cal M}_n=\{n\}$ and $W_m=1$ for all $m\in{\cal M}$. The energy and bandwidth allocation subproblem can be rewritten as
 \begin{equation}
{\sf EP}_n:\quad \max_{\boldsymbol{p}_n}  \sum_{k=1}^K  a_n^k\cdot \log(1+\frac{p_n^kH_n^k}{a_n^k})
\end{equation}
subject to
\begin{equation}
\left\{
\begin{array}{l}
\tilde{E}_n^k - B_n^{\max} \leq  \sum_{\kappa=1}^{k}p_n^{\kappa} \leq \tilde{E}_n^k \\
0 \leq p_n^k\leq P_n\\
\end{array}\right. ,\ k\in{\cal K}\ ,
\end{equation}
and
\begin{equation}
{\sf BP}_k(\epsilon):\quad\max_{\boldsymbol{a}^k}\sum_{n=1}^N a_n^k\cdot \log(1+\frac{p_n^kH_n^k}{a_n^k})
\end{equation}
subject to
\begin{equation}
\left\{
\begin{array}{l}
\sum_{i=1}^N a_i^k = 1\\
a_n^k \geq \epsilon,\ n\in{\cal N}
\end{array}\right. \ ,
\end{equation}
respectively.
 

\subsection{Solving {\sf EP}$_n$: Discounted Dynamic Water-Filling}
Given $\boldsymbol{a}^k$, since EP$_n$ is a subproblem of the problem in \eqref{eq:aproblem}-\eqref{eq:acst}, its K.K.T. conditions form a subset of those of the original problem, given by \eqref{kkt:wf}, \eqref{kkt:a} and \eqref{kkt:b}.

To develop an efficient algorithm, we first rewrite the K.K.T. condition in \eqref{kkt:wf} as
\begin{equation}\label{eq:wf0}
p_n^k = a_n^k\cdot \Big(\frac{1}{v_n^k - u_n^k} - \frac{1}{H_n^k}\Big)\ .
\end{equation}
Since the energy allocation must satisfy $0\leq p_n^k \leq P_n$,  \eqref{eq:wf0} can be further written as
\begin{equation}\label{eq:wf}
p_n^k = \min\Big\{P_n, a_n^k\cdot \Big[\frac{1}{v_n^k - u_n^k} - \frac{1}{H_n^k}\Big]^+\Big\}\ .
\end{equation}

Comparing the K.K.T. conditions in \eqref{kkt:wf} with (24) in \cite{2014arXiv1401.2376W}, the only difference is the scaling factor $a_n^k$. Following the same analysis in \cite{2014arXiv1401.2376W}, we have the following theorem.

%

\begin{theorem}\label{thm:wf}
Given any bandwidth allocation $\cal A$, a feasible energy allocation $\cal P$ is an optimal solution to \eqref{eq:problem}-\eqref{eq:cst}, if and only if it follows the discounted water-filling rule in \eqref{eq:wf}, where the water level $\frac{1}{v^k-u^k}$ may increase only at a battery depletion point (BDP) such that $B_n^k = 0$,  and decrease only at a battery fully charged point (BFP) such that $B_n^k=B_n^{\max}$.
\end{theorem}

Theorem \ref{thm:wf} gives the necessary and sufficient conditions for the optimal energy allocation given any bandwidth allocation. Given the set of BDP/BFPs corresponding to the optimal energy allocation, the optimal energy allocation and the corresponding the water level for the segment between two adjacent BDP/BFPs, $(a,\textrm{type of } a)$ and $(b,\textrm{type of } b)$, can be written as
\begin{equation}\label{eq:swfc}
p_n^k = \min\Big(P_n\;,\;a_n^k\cdot\big[w- \frac{1}{H_n^k}\big]^+\Big)\ ,
\end{equation}
where $w$ is the {\em water level of a segment} such that
\begin{equation}\label{eq:waterlevel}
\sum_{\kappa=a+1}^b p_n^{\kappa} = \min\Big\{(b-a)P_n,{\tilde E}_n^b-{\tilde E}_n^{a} + \big(\mathbb{I}(a\textrm{ is BFP})-\mathbb{I}(b\textrm{ is BFP})\big)B_n^{\max}\Big\}\ ,
\end{equation}
with $\mathbb{I}(\cal A)$ being an indicator function given by
\begin{equation}\label{eq:idc}
\mathbb{I}({\cal A})\triangleq
\left\{\begin{array}{ll}
1,&\textrm{if } {\cal A} \textrm{ is true}\\
0,&\textrm{otherwise}\\
\end{array}\right.\ .
\end{equation}
Specifically, \eqref{eq:swfc}-\eqref{eq:waterlevel} represent the water-filling operation in a segment between two optimal BDP/BFPs, as mentioned in Theorem \ref{thm:wf}. Also, \eqref{eq:waterlevel} ensures that with the energy allocation the boundary points $a$ and $b$ are the desired BDP/BFPs.

In \cite{2014arXiv1401.2376W},  to obtain the optimal energy schedule for a single energy harvesting transmitter, a dynamic water-filling algorithm is proposed that recursively performs the ``forward search''  and ``backward search'' operations on a water-filling attempting basis, i.e., in the above two operations we perform conventional water-filling on each ``target'' segment (some consecutive slots).  Using the dynamic water-filling algorithm, the set of the optimal BDP/BFPs can be obtained and the corresponding water level between two adjacent BDP/BFPs satisfies the optimality conditions for the energy scheduling problem. Then, it is easy to verify that, by replacing the conventional water-filling in the dynamic water-filling algorithm by the water-filling operations defined in \eqref{eq:swfc}-\eqref{eq:waterlevel}, we can still obtain the optimal BDP/BFPs set for the energy allocation subproblem and thus the optimal energy allocation can be finally obtained. We call the new algorithm the {\em discounted dynamic water-filling algorithm}, whose computational complexity is ${\cal O}(K^2)$ \cite{2014arXiv1401.2376W}.


%
%
\subsection{Solving {\sf BP}$_k(\epsilon)$:  Bandwidth Fitting Algorithm}
We first note that, when $p_n^k=0$ and $a_n^k \geq \epsilon$, the channel rate achieved by transmitter $n$ in slot $k$ is zero. Therefore, in a slot $k$ such that $\sum_{n=1}^N p_n^k = 0$, any feasible bandwidth allocation is optimal, achieving the maximum channel rate $0$. However, in a slot $k$ where $\sum_{n=1}^N p_n^k>0$, in order to maximize the channel rate, the transmitter with zero energy allocation $p_n^k=0$ must be allocated with the minimal bandwidth, i.e., $a_n^k = \epsilon$. We denote the energy allocation $\{p_n^k\;|\;\sum_{i=1}^N p_i^k > 0, n\in{\cal N}\}$ as the {\it non-zero energy allocation} and, in the remainder of this subsection, we will obtain the optimal bandwidth allocation given a non-zero energy allocation.

Since {\sf BP}$_k(\epsilon)$ is a subproblem of the problem given in \eqref{eq:aproblem}-\eqref{eq:acst}, its K.K.T. conditions form a subset of those of the original problem, given by \eqref{kkt:al}, \eqref{kkt:c} and \eqref{kkt:d}.

Given a non-zero energy allocation, since we know that the transmitter with zero energy allocation should be allocated with the minimal bandwidth, we rewrite the K.K.T. condition related to the transmitters with the non-zero energy allocation  in \eqref{kkt:al} as
\begin{equation}\label{eq:equ0}
-\log(\frac{a_n^k}{a_n^k + p_n^kH_n^k})-(1-\frac{a_n^k}{a_n^k + p_n^kH_n^k})= \alpha^k-\beta_n^k
\end{equation}
for all $k\in{\cal K}$, and all $n$ such that $p_n^k >0, n\in{\cal N}$.

Denoting $y(\alpha^k,\beta_n^k) $ as the solution to the following equation,
\begin{equation}\label{eq:equ}
-\log(y(\alpha^k,\beta_n^k) ) - (1-y(\alpha^k,\beta_n^k) ) = \alpha^k - \beta_n^k\\ ,
\end{equation}
from \eqref{eq:equ0}-\eqref{eq:equ}, we then have
\begin{equation}\label{eq:al0}
a_n^k = p_n^kH_n^k\cdot z(\alpha^k,\beta_n^k)\ ,
\end{equation}
where
\begin{equation}
z(\alpha^k,\beta_n^k) \triangleq \frac{y(\alpha^k,\beta_n^k) }{1-y(\alpha^k,\beta_n^k)}\ ,
\end{equation}
and $0 < y(\alpha^k,\beta_n^k)< 1$.

By \eqref{kkt:d}, we have that $\beta_n^k$ must be zero when $a_n^k>\epsilon$. Then we divide $\{a_n^k,n\in{\cal N}\}$ into two sets: ${\cal T}^k\triangleq\{n\;|\;a_n^k >\epsilon,n\in{\cal N}\}$ and $\bar{\cal T}^k\triangleq\{n\;|\;a_n^k =\epsilon,n\in{\cal N}\}$. Then, \eqref{eq:al0} can be rewritten as
\begin{equation}\label{eq:al3}
a_n^k=
\left\{
\begin{array}{ll}
p_n^kH_n^k z(\alpha^k,0),&\textrm{ when }n\in{\cal T}^k, p_n^k > 0\\
p_n^kH_n^k z(\alpha^k,\beta_n^k) ,&\textrm{ when }n\in\bar{T}^k,p_n^k > 0\\
\end{array}\right. ,
\end{equation}
where $\beta_n^k \geq 0$ and
\begin{equation}\label{eq:al2}
p_n^kH_n^k z(\alpha^k,{\beta}_n^k) = \epsilon, \ n\in\bar{\cal T}^k\ .
\end{equation}
Also, we have
\begin{equation}\label{eq:al00}
a_n^k = \epsilon \geq p_n^kH_n^k z(\alpha^k,{\beta}_n^k) ,\ \textrm{when }p_n^k = 0 \ .
\end{equation}

Substituting \eqref{eq:al3} and \eqref{eq:al00} into the constraints $\sum_{n=1}^N a_n^k=1$, we further have
\begin{equation}\label{eq:al1}
|\bar{\cal T}^k|\cdot \epsilon + \sum_{n\in{\cal T}^k} p_n^kH_n^k z(\alpha^k,0) = 1\ .
\end{equation}


%

%

Next, we characterize the optimality condition for the bandwidth allocation problem BP$_k(\epsilon)$ given the non-zero energy allocation, as follows.
\begin{theorem}\label{thm:al}
Given any energy allocation $\{p_n^k\;|\;\sum_{i=1}^Np_i^k > 0, n\in{\cal N}\}$, the bandwidth allocation $\boldsymbol{a}^k$ is the optimal to ${\sf BP}(\epsilon)$, if and only if, for every $n\in {\cal T}^k$, it satisfies
\begin{equation}\label{eq:bal1}
a_n^k = (1-|\bar{\cal T}^k|\cdot \epsilon) \frac{p_n^kH_n^k}{\sum_{i\in{\cal T}^k}p_i^kH_i^k},
\end{equation}
and for any $n\in\bar{\cal T}^k$, it satisfies
\begin{align}
a_n^k=\epsilon \geq   (1-|\bar{\cal T}^k|\cdot \epsilon) \frac{p_n^kH_n^k}{\sum_{i\in{\cal T}^k}p_i^kH_i^k}\ ,\label{eq:bal2}
\end{align}
for all $k\in{\cal K}$.
\end{theorem}
\begin{IEEEproof}
Rearranging \eqref{eq:al1}, we have
\begin{equation}\label{eq:ral1}
z(\alpha^k,0) = \frac{1 - |\bar{\cal T}^k|\cdot \epsilon}{\sum_{n\in{\cal T}^k} p_n^kH_n^k }\ .
\end{equation}

Necessity: When $a_n^k\in{\cal T}^k$, we have $a_n^k>\epsilon$ and thus $\beta_n^k=0$. Substituting \eqref{eq:ral1} to \eqref{eq:al3}, we have \eqref{eq:bal1}. When $a_n^k\in\bar{\cal T}^k$,  we have $a_n^k=\epsilon$ and thus $\beta_n^k\geq0$.

Note that, since $y(\alpha^k,\beta_n^k)$ is the solution to \eqref{eq:equ}, which is an equation of the form $y - \log(y)=x$ for which $y$ increases when $x$ decreases for $y\in(0,1)$, we see that $y(\alpha^k,\beta_n^k)$ increases as $\beta_n^k$ increases. Then, we have $z(\alpha^k,\beta_n^k)$  increases as $\beta_n^k$ increases given $\alpha^k$, thus $z(\alpha^k,\beta_n^k)\geq z(\alpha^k,0)$. Since $p_n^kH_n^k\geq 0$, we further have
\begin{equation}\label{eq:ineq}
p_n^kH_n^kz(\alpha^k,\beta_n^k)\geq p_n^kH_n^kz(\alpha^k,0)\ .
\end{equation}
Substituting \eqref{eq:al2} or \eqref{eq:al00}, and \eqref{eq:ral1} into the LHS and RHS of \eqref{eq:ineq}, respectively, we get \eqref{eq:bal2}.

Sufficiency: For the transmitters with zero energy allocation, it is easy to verify that the minimal bandwidth allocation is optimal. For other transmitters, since $a_n^k>0$, by \eqref{eq:al0}, we must have $0< y(\alpha^k,\beta_n^k) < 1$ and thus $0<z(\alpha^k,\beta_n^k)<\infty$. Note that, by \eqref{eq:equ}, $ y(\alpha^k,\beta_n^k)\in(0,1)$ and $\alpha^k-\beta_n$ is one-to-one mapping. Therefore, for any $a_n^k$ satisfying the sufficient conditions in Theorem \ref{thm:al}, we can always find the corresponding dual variables $\alpha^k$ and $\beta_n^k$ in \eqref{eq:al1} and \eqref{eq:al2}, satisfying all the K.K.T. conditions.
\end{IEEEproof}

Intuitively, Theorem \ref{thm:al} states that the optimal bandwidth allocation should be proportional to the transmission ``condition'', i.e., $p_n^kH_n^k$. In particular, if the desired bandwidth allocation for transmitter $n$ is less than the minimal requirement $\epsilon$, $a_n^k$ should be set as the minimal requirement $\epsilon$. Based on Theorem \ref{thm:al}, we propose the following {\em iterative bandwidth fitting algorithm}. Initially, we set ${\cal T}^k=\{n\;|\;p_n^k>0, n\in{\cal N} \}$ and $\bar{\cal T}^k={\cal N}\setminus {\cal T}^k$. In each iteration, we calculate the bandwidth allocation $a_n^k$ by \eqref{eq:bal1} with the current ${\cal T}^k$ and $\bar{\cal T}^k$. We denote ${\cal V}=\{n\;|\;a_n^k\leq \epsilon,n\in{\cal T}^k\}$ as a ``violation set'', containing the elements in ${\cal T}^k$ that violate the definition of ${\cal T}^k\triangleq\{n\;|\;a_n^k >\epsilon,n\in{\cal N}\}$.  Then, we move all $n\in{\cal V}$ from  ${\cal T}^k$ to $\bar{\cal T}^k$. This iterative process ends when ${\cal V}$ is empty. Finally, with the obtained  ${\cal T}^k$ and $\bar{\cal T}^k$, the optimal allocation can be calculated by \eqref{eq:bal1}-\eqref{eq:bal2}.

The procedure of the algorithm is summarized as follows: \\
\begin{minipage}[h]{6.5 in}
\rule{\linewidth}{0.3mm}\vspace{-.1in}
{\bf {\footnotesize Algorithm 2 - Iterative Bandwidth Fitting Algorithm}}\vspace{-.2in}\\
\rule{\linewidth}{0.2mm}
{ {\small
\begin{tabular}{ll}
	\;1:&  Initialization\\
	\;& ${\cal T}^k=\{n\;|\; p_n^k > 0, n\in{\cal N}\}$, $\bar{\cal T}^k={\cal N}\setminus {\cal T}^k$\\
    \;2:&Bandwidth Fitting\\
    	\;&{\bf FOR} $k\in{\cal K}$ such that $\sum_{n=1}^N p_n^k > 0$\\
    	\;&\quad {\bf REPEAT}\\
    	\;& \quad \quad Calculate $a_n^k$ by \eqref{eq:bal1} for all $n\in {\cal T}^k$\\
    	\;& \quad \quad Set the violation set ${\cal V}=\{n\;|\;a_n^k\leq \epsilon,n\in{\cal T}^k\}$\\
    	\;&\quad\quad Move all $n\in\cal V$ from  ${\cal T}^k$ to $\bar{\cal T}^k$\\
    	\;&\quad{\bf UNTIL} ${\cal V}=\{\ \}$\\
      \;& {\bf ENDFOR}\\
    \;2:& Bandwidth Allocation\\
    \;& Obtain $a_n^k$ with ${\cal T}^k$ and $\bar{\cal T}^k$ by \eqref{eq:bal1}-\eqref{eq:bal2} for all $n\in{\cal N},k\in{\cal K}$ such that $\sum_{n=1}^N p_n^k > 0$
\end{tabular}}}\\
\rule{\linewidth}{0.3mm}
\end{minipage}\vspace{.1 in}\\

Note that, Algorithm 2 will terminate in at most $N$ iterations since the elements transfer between  ${\cal T}^k$ and $\bar{\cal T}^k$ is one-directional. Moreover, for all $k\in{\cal K}$ such that $\sum_{n=1}^N p_n^k > 0$, at the end of the last iteration, since  ${\cal V}$ is empty, we have that the condition in \eqref{eq:bal1} is satisfied by  all $n\in{\cal T}^k$ and  we have $a_n^k>\epsilon$ for all $n\in{\cal T}^k$. Also, for all other $n\in\bar{\cal T}^k$, obviously we have $a_n^k=\epsilon$. If \eqref{eq:bal2} is also satisfied, we can further claim that with the obtained  ${\cal T}^k$ and $\bar{\cal T}^k$, Algorithm 2 gives an optimal bandwidth allocation.

The next result shows that, at the end of each iteration (including the last iteration), with the obtained $\bar{\cal T}^k$, \eqref{eq:bal2} is satisfied. The proof is given in Appendix B.

\begin{proposition}\label{pp:al}
For all $n\in\bar{\cal T}^k$, which is obtained by Algorithm 2 at the end of each iteration, \eqref{eq:bal2} is satisfied for all $k\in{\cal K}$ such that $\sum_{n=1}^N p_n^k > 0$.
\end{proposition}

With Proposition \ref{pp:al}, we conclude that the bandwidth allocation obtained by Algorithm 2 is optimal. Moreover, since the number of iterations is bounded by $N$, the computational complexity of Algorithm 2 is ${\cal O}(N)$.

\begin{remark}
In each iteration of Algorithm 1, $N$ subproblems of {\sf EP}$_n$ and $K$ subproblems of {\sf BP}$_k(\epsilon)$ need to be solved, using the discounted dynamic water-filling algorithm and the bandwidth fitting algorithm, whose computational complexities are ${\cal O}(K^2)$ and ${\cal O}(N)$, respectively. Thus the overall computational complexity of Algorithm 1 becomes ${\cal O}(NK^2)$, which is significantly lower than that of the generic convex tools.
\end{remark}

\section{Suboptimal Algorithm with Causal Information}
In Section III, we proposed an iterative algorithm to obtain the optimal energy-bandwidth allocation with non-causal information of the channel gains and the harvested energy, whose performance can also serve as an upper bound on the achievable rate. In this section, we consider the case that the channel fading and energy harvesting are not predicable, i.e., their realizations can only be observed causally at the beginning of the corresponding slot. We will propose a heuristic algorithm to obtain the suboptimal energy-bandwidth allocation that follows the structure of the optimal solution. For simplicity, we still focus on the throughput maximization problem for point-to-point channels considered in Section IV.

We first give the structure of the optimal solution for the problem in \eqref{eq:problem}-\eqref{eq:cst}.
\begin{lemma}\label{lm:nec}
If $({\cal A},{\cal P})$ is the optimal solution to the problem in \eqref{eq:problem}-\eqref{eq:cst}, then
\begin{itemize}
\item $\{p_n^k\;|\;a_n^k > 0, n\in{\cal N}, k\in{\cal K}\}$ satisfy
\begin{equation}\label{eq:lwf}
p_n^k = \min\Big\{P_n, a_n^k\cdot \Big[w_n^k - \frac{1}{H_n^k}\Big]^++\gamma_n^k\Big\}\ ,
\end{equation}
where $\gamma_n^k$ is the energy adjuster and $w_n^k>0$ may only increase/decrease at BDP/BFP;
\item $\{a_n^k\;|\; \sum_{i=1}^N p_i^k>0, n\in{\cal N}, k\in{\cal K}\}$ satisfy
\begin{equation}\label{eq:lba}
a_n^k = \frac{p_n^kH_n^k}{\sum_{i\in{\cal N}}p_i^kH_i^k} \ .
\end{equation}
\end{itemize}
\end{lemma}
\begin{IEEEproof}
Comparing \eqref{eq:lwf} with the optimal energy allocation given in \eqref{eq:wf} by Theorem \ref{thm:wf}, the only difference is the term of energy adjuster $\gamma_n^k$. Note that, for the optimal energy allocation, given $n,k$, we have that $\gamma_n^k =0$ if $p_n^k < P_n$ since the performance can be improved if $p_n^k$ can be further increased by decreasing a positive $\gamma_n^k$. Therefore, we have that for the optimal energy allocation,  \eqref{eq:lwf} is equivalent to \eqref{eq:wf}.

Given $\cal P$, we next show that the optimal $a_n^k=0$ if and only if $p_n^k=0$. Specifically, if $p_n^k=0$, the rate of link $n$ in slot $k$ is constant zero for any $a_n^k \geq 0$. Since $\sum_{i=1}^N p_i^k>0$ and $\sum_{n=1}^N a_n^k = 1$, if we reassign the non-zero bandwidth $a_n^k$ of link $n$ to any other links $i$ such that $p_i^k > 0$ in slot $k$, the sum rate in slot $k$ is increased. Therefore, we have $a_n^k=0$ if $p_n^k=0$. On the other hand, if $p_n^k > 0$, we have that the derivative of the objective function over $a_n^k$ tends to infinity as $a_n^k \rightarrow 0^+$, which means that we can always move certain bandwidth from some other link to link $n$ with $a_n^k = 0$ and $p_n^k > 0$, such that the rate loss of the other link is less than the rate gain of link $n$. Therefore, $a_n^k=0$ is not optimal if $p_n^k > 0$. Hence, $a_n^k=0$ if and only if $p_n^k=0$. By eliminating the terms of $p_n^k=0$ in the objective function, the optimal bandwidth allocation is positive and then Theorem \ref{thm:al} can be adapted for the case $\epsilon=0$, i.e., \eqref{eq:lba} is obtained.
\end{IEEEproof}

Lemma \ref{lm:nec} provides the structure of the optimal energy-bandwidth allocation, in which the water level $w_n^k$ is the only parameter affected by the future channel fading and energy harvesting. Specifically, if the energy harvesting and channel gains are predictable, then the optimal $w_n^k$ can be obtained, as in the proposed non-causal algorithm, where $\gamma_n^k=0$ if $p_n^k < P_n$, and $\gamma_n^k \geq 0$ if $p_n^k = P_n$. In other words, in \eqref{eq:lwf}, $\gamma_n^k$ does not affect the value of $p_n^k$ when the optimal water level $w_n^k$ is given. However, when the energy harvesting and channel fading  processes  are unpredictable, the optimal $w_n^k$ is hard to obtain. Note that $\gamma_n^k$ essentially acts as an adjusting factor to mitigate the energy waste caused by the non-optimality of $w_n^k$, i.e., if the suboptimal water level is lower than the optimal one and therefore causes the energy waste, we can try to utilize the wasted energy for transmission. Then, we use the potentially wasted energy as the adjuster, given by
\begin{equation}\label{eq:adjst}
 \gamma_n^k  = \max\left\{0,B_n^{k-1} +\Delta_n^k- \min\Big\{P_n, a_n^k\cdot \Big[w_n^k - \frac{1}{H_n^k}\Big]^+\Big\}-B_n^{\max}\right\}\ .
\end{equation}
where  $\Delta_n^k \triangleq E_n^k - E_n^{k-1}$ is the energy harvested energy in slot $k$. Specifically, $\gamma_n^k$ becomes the actual energy wastage $D_n^k$ if the water-filling fashion in \eqref{eq:wf} is followed using the water level $w_n^k$.

Based on Lemma \ref{lm:nec}, we design an {\em adaptive water-filling algorithm}, aiming to obtain a suboptimal energy-bandwidth allocation, which follows the structure of the optimal solution given in Lemma \ref{lm:nec}. With the proposed algorithm, except for the calculation of the water levels, all other optimality conditions are approached by the obtained energy-bandwidth allocation. Specifically, to avoid the use of the future information, the water levels are calculated by a heuristic method.


The proposed algorithm is an online algorithm. Initially, we set a water level $w_n^0$ for each transmitter $n\in{\cal N}$. At the beginning of slot $k$, we check the battery level of each transmitter. If the battery is empty or full, we decrease or increase the water level by a factor, e.g., $w_n^k= c\cdot w_n^{k-1} (\textrm{or } w_n^{k-1}/c)$. Otherwise, we keep the water level unchanged. Then, based on the water level $w_n^k$, we calculate the energy allocation and bandwidth allocation $\{p_n^k,a_n^k\;|\;n\in{\cal N}\}$ by solving the equations \eqref{eq:lwf}, \eqref{eq:lba} and \eqref{eq:adjst}.
In particular, substituting \eqref{eq:adjst} into \eqref{eq:lwf}, there are two equations and two variables, which can be solved numerically.

Moreover, we propose the following choices of the initial water level $w_n^0$ and the factor $c$,
\begin{equation}
w_n^0 \approx N\cdot\mathbb{E}\left[{E_n^k}\right] + \mathbb{E}\left[\frac{1}{H_n^k}\right]\ ,
\end{equation}
and
\begin{equation}
c \approx 1 + P_n /  w_n^0
\end{equation}

The algorithm is summarized as follows: \\
\begin{minipage}[h]{6.5 in}
\rule{\linewidth}{0.3mm}\vspace{-.1in}
{\bf {\footnotesize Algorithm 3 - Adaptive Water-Filling Algorithm (the superscript $k$ is dropped)}}\vspace{-.2in}\\
\rule{\linewidth}{0.2mm}
{ {\small
\begin{tabular}{ll}
	\;1:&  Input\\
	\;& Current water level and battery level $\{w_n,B_n\;|\;n\in{\cal N}\}$\\
    	\;2:& Output \\
	\;& Updated  water level and battery level $\{w_n,B_n\;|\;n\in{\cal N}\}$\\
	\;& Energy-bandwidth allocation $\{p_n,a_n\;|\;n\in{\cal N}\}$\\
	\;3:& At the beginning of each slot\\
    	\;&{\bf FOR} $n\in{\cal N}$\\
    	\;&\quad {\bf IF} $B_n=B_n^{\max}$ {\bf THEN} $w_n \leftarrow w_n/c$\\
    	\;&\quad {\bf IF} $B_n=0$ {\bf THEN} $w_n \leftarrow w_n\cdot c$\\
      \;& {\bf ENDFOR}\\
      \;& Solve the equation group of \eqref{eq:lwf}, \eqref{eq:lba} and \eqref{eq:adjst} to obtain $\{p_n,a_n\;|\;n\in{\cal N}\}$\\
      \;& $B_n \leftarrow \min\{P_n, B_n + \Delta_n - p_n\}$ for all $n\in{\cal N}$
\end{tabular}}}\\
\rule{\linewidth}{0.3mm}
\end{minipage}\vspace{.2 in}\\

\section{Simulation Results}
Suppose that there are $N=4$ transmitters in the network and each communicates with one receiver, and we assume $W_n=1$ for $n=1,2,3,4$. We set the scheduling period as $K=40$ slots. For each transmitter $n$, we set the initial battery level $B_n^0=0$ and the maximum battery capacity $B_n^{\max} = 20$ units. Assume that the harvested energy $E_n^k$ follows a truncated Gaussian distribution with mean $\mu_E$ and variance of $\sigma^2=2$, and the fading channel  parameter follows the standard complex Gaussian distribution, i.e., $h_n^k \sim {\cal CN} (0,1)$,
so that $H_n^k\sim \exp(1)$.

For comparison, we consider three scheduling strategies, namely, the {\em greedy policy}, the {\em TDMA greedy policy}, and the {\em equal bandwidth policy}. For the greedy policy, each transmitter tries to consume the harvested energy as much as possible in each slot, as calculated by \eqref{eq:greedy}. Then, the central controller allocates the bandwidth to each transmitter by using the iterative bandwidth fitting algorithm (i.e., Algorithm 2). For the TDMA greedy policy, each transmitter uses the maximum possible energy to transmit in each slot, and the central controller allocates the entire bandwidth to the transmitter with the maximum $p_n^kH_n^k$. For the equal bandwidth policy, the central controller allocates each transmitter equal bandwidth and then each transmitter uses the optimal energy allocation.

To evaluate the performance of different  algorithms, we consider two scenarios, namely, the {\em energy-limited} scenario, where the maximum transmission power is $P_n = 10$ units per slot, and the {\em power-limited} scenario, where the maximum transmission power is $P_n = 5$ units per slot. Moreover, the convergence threshold in Algorithm 1 is set as $\delta = 10^{-3}$, and the initial water level  and the parameter $c$ in Algorithm 3 are set as $w_n^0=25$ and  $c = 1.1$, respectively. We compare the achievable rates of different algorithms under different mean values $\mu _E$  of the harvested energy.  In the energy-limited scenario, the transmitter has more freedom to schedule the harvested energy to be consumed in each slot because of  the large maximum transmission power. One the other hand, in the power-limited scenario, the harvested energy would be consumed in the future slots since the maximum transmission power is reached more frequently. Furthermore, for both scenarios, when the energy harvesting parameter $\mu_E$ is small, it corresponds to the ``energy-constrained'' condition, where the scheduling is mainly constrained by the energy availability. And when $\mu_E$ is large, it corresponds to the ``power-constrained'' condition, where the scheduling is more constrained by the maximum transmission power.

\begin{figure}
\centering
\includegraphics[width=0.6\textwidth]{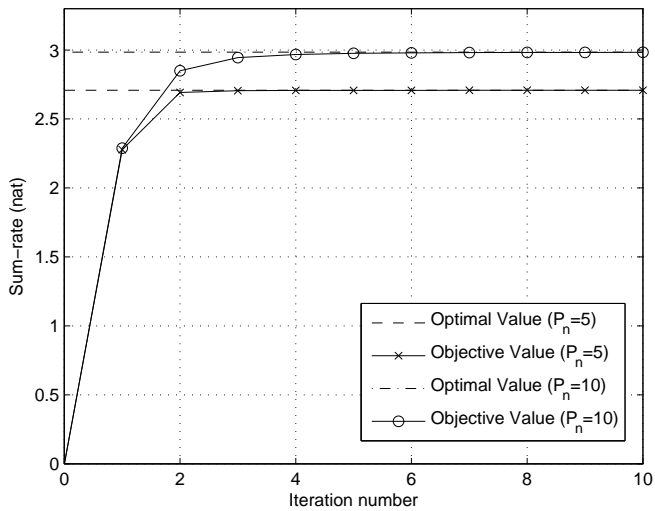}
\caption{The convergence of Algorithm 1 for $\mu_E=4$.}
\label{fg:3}
\end{figure}

\begin{figure}
\centering
\includegraphics[width=0.6\textwidth]{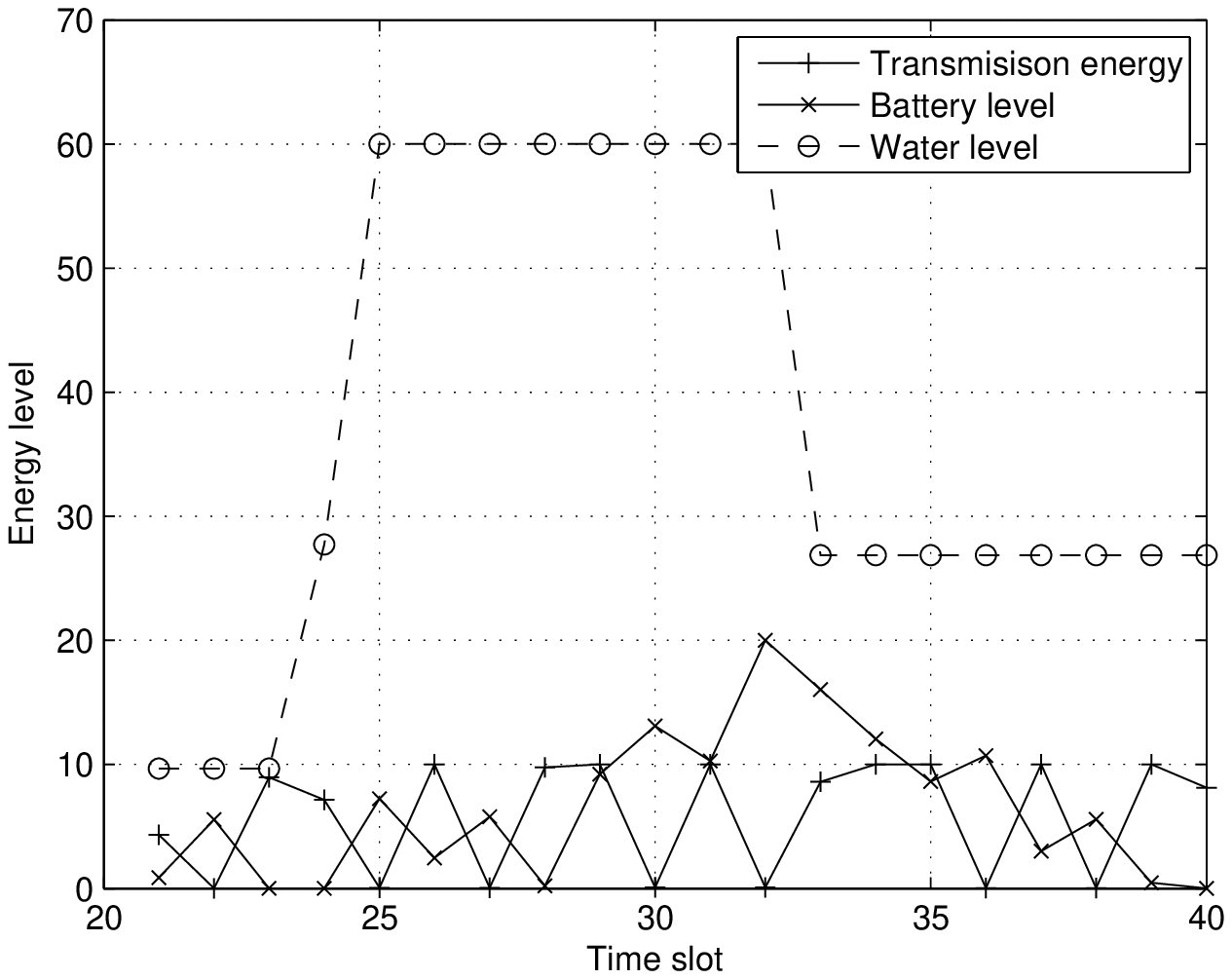}
\caption{20-slot snapshot of the optimal energy allocation obtained by Algorithm 1 for a particular transmitter ($\mu_E=4,P_n=10$).}
\label{fg:21}
\end{figure}

\begin{figure}
\centering
\includegraphics[width=0.6\textwidth]{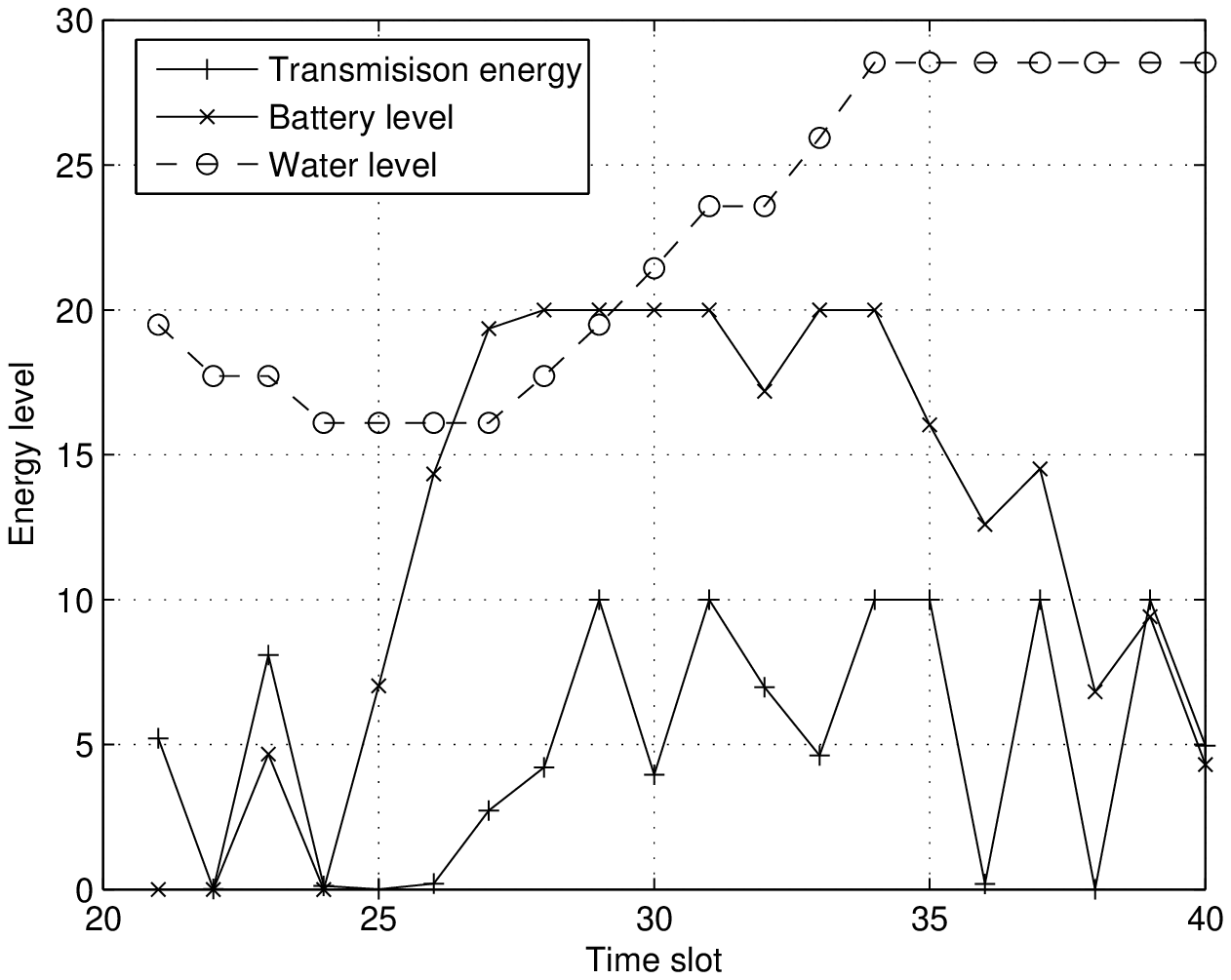}
\caption{20-slot snapshot of the energy allocation obtained by Algorithm 3 for a particular transmitter ($\mu_E=4,P_n=10$).}
\label{fg:22}
\end{figure}

\begin{figure}
\centering
\includegraphics[width=0.6\textwidth]{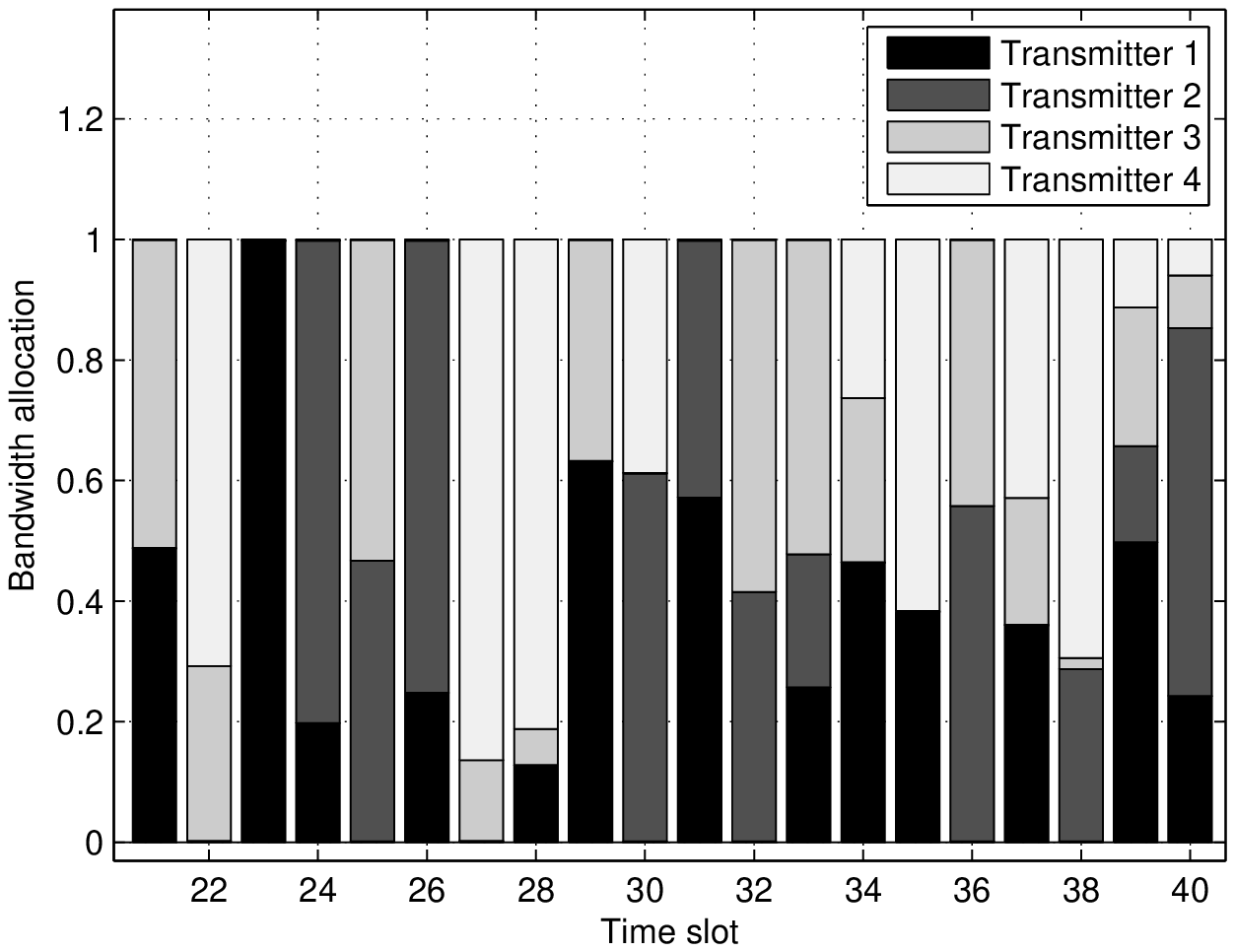}
\caption{20-slot snapshot of the optimal bandwidth allocation by Algorithm 1 ($\mu_E=4,P_n=10$).}
\label{fg:23}
\end{figure}

Before we compare the performance of different algorithms, we first illustrate the convergence behavior of Algorithm 1 in Fig. \ref{fg:3} with $\mu_E=4$. It is seen that Algorithm 1 converges  (the relative error is less than $0.001$) within $4$ and $7$ iterations for $P_n=5$ and $P_n=10$, respectively. Next, we set  $\mu_E = 4$ and $P_n = 10$ and give the $20$-slot snapshots (slot 20 - slot 40) of the obtained energy-bandwidth allocation in Fig. \ref{fg:21}, Fig. \ref{fg:22} and Fig. \ref{fg:23}. Specifically, Fig. \ref{fg:21} and Fig. \ref{fg:22} illustrate the relationship among the water level $w_n^k$, transmission energy $p_n^k$ and the battery level $B_n^k$ obtained by Algorithm 1 and Algorithm 3, respectively, showing that although both algorithms follow the water-filling structure with the dynamic water levels, their water levels vary according to different rules, based on the dynamic of the battery. Moreover,  the optimal bandwidth allocation $a_n^k$ obtained by Algorithm 1 is illustrated in Fig. \ref{fg:23}, and we can see that most of the time the channel is shared by multiple transmitters to maximize the sum-rate.


\begin{figure}
\centering
\includegraphics[width=0.6\textwidth]{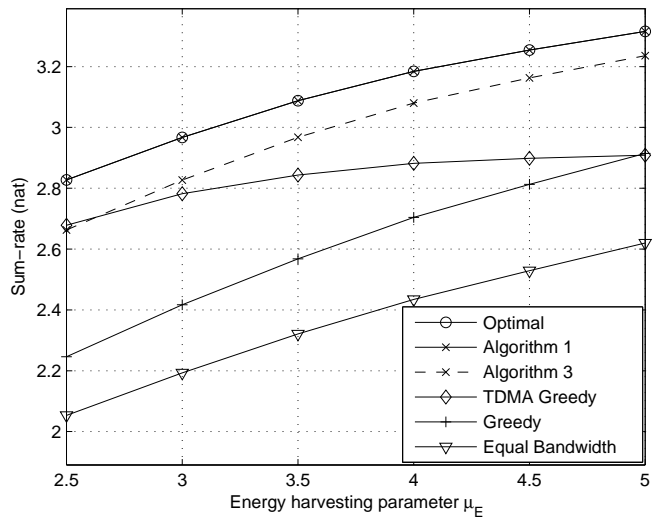}
\caption{Performance comparisons in the energy-limited scenario ($P_n=10$, $B_n^{\max}=20$).}
\label{fg:1}
\end{figure}

\begin{figure}
\centering
\includegraphics[width=0.6\textwidth]{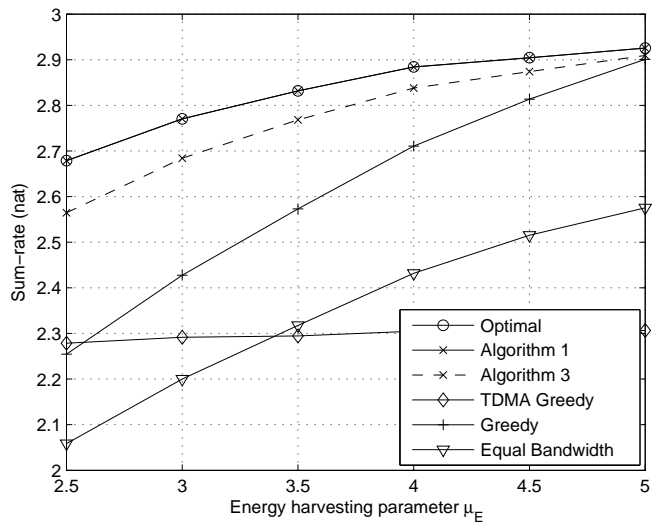}
\caption{Performance comparisons in the power-limited scenario ($P_n=5$, $B_n^{\max}=20$).}
\label{fg:2}
\end{figure}

We then run the simulation $1000$ times to obtain the rates given by various scheduling strategies, as well as by the optimal schedule solved  by a general convex solver, shown in Fig. \ref{fg:1} and Fig. \ref{fg:2}, for the energy-limited scenario and the power-limited scenario, respectively.  It is seen from that for both scenarios the proposed non-causal iterative algorithm (Algorithm 1) achieves the same performance as that corresponding to the optimal energy-bandwidth allocation solved by the generic convex solver, corroborating the optimality of Algorithm 1 as stated by Theorem 2. Also, the proposed causal algorithm (Algorithm 3) performs worse than the optimal policy but still better than the other heuristic policies. Moreover, for all policies, the performance is improved as the mean of the harvested energy increases.

From Fig. \ref{fg:1}, for the  energy-limited scenario, the performance gap between the TDMA greedy policy and the optimal solution increases as the mean of the harvested energy increases. It is because when the mean of the harvested energy is high, due to the maximum transmission power and battery capacity constraints, the single-user transmission of TDMA results in significant energy waste by the non-transmitting transmitters in each slot.

On the other hand, from Fig. \ref{fg:2},  for the power-limited scenario, the performance  gap between the optimal solution and some of the suboptimal algorithms (Algorithm 3 and the greedy policy) decreases as the mean of the harvested energy increases. It is because when the harvested energy is ample, the optimal energy allocation achieves the maximum transmission power more frequently and approaches  the greedy policy. Also, in the  power-limited scenario, the TDMA greedy policy performs significantly worse than other algorithms since the low maximum transmission power results in a lot of energy waste in the absence of any channel sharing.

Moreover, as expected, the performance in the energy-limited scenario is better than that in the power-limited energy for all policies. This is because the lower maximum transmission power restricts the flexibility of the energy scheduling and causes waste  of energy due to the limited battery capacity.

\section{Conclusions}

In this paper, we have considered the joint energy-bandwidth allocation problem for multiple energy harvesting transmitters over $K$ time slots. This problem is formulated as a convex optimization problem with ${\cal O}(MK)$ variables and constrains, where $M$ is the number of the receivers and $K$ is number of the slots in a scheduling period, which is hard to solve with a generic convex tool. We have proposed an  energy-bandwidth allocation algorithm that iterates between solving the energy allocation subproblem and the bandwidth allocation subproblem, and the convergence and the optimality of the iterative algorithm have been shown. When each transmitter communicates with one receiver and the sum-rate is unweighted, the discounted dynamic water-filling algorithm and the bandwidth fitting algorithm are proposed to optimally solve the energy and bandwidth allocation subproblems, respectively. Moreover, a heuristic algorithm is also proposed to obtain the suboptimal energy-bandwidth allocation causally and efficiently, by following the structure of the optimal energy-bandwidth solution.  In a companion paper, we will consider multiple broadcast channels under the joint energy-bandwidth allocation framework and develop efficient algorithms for solving the two subproblems under both orthogonal and non-orthogonal access.


%

\appendices
\section{Proof of Theorem \ref{pp:cvg}}

We note that, the feasible domain of ${\sf BP}_{k}(\epsilon_0 /i)$ expands with iterations while  the feasible domain of ${\sf EP}_n$ remains unchanged. Since we successively solve the maximization problems {\sf EP}$_n$ and {\sf BP}$_k(\epsilon_0/i)$ in iteration $i$, we have that the objective value is non-decreasing over the iterations. On the other hand, the objective function is upper bounded by $C_{\cal W}({\cal P,A})\leq \sum_{m=1}^M\sum_{k=1}^K\log(1+P_mH_m^k)$ therefore the algorithm converges.
Since the feasible domain of ${\sf P}_{\cal W}(\epsilon)$ is a closed set for $\epsilon \geq 0$, at the converged point $V$, we can find the corresponding ${\cal P}_0$ and ${\cal A}_0$ which are pairwise optimal for ${\sf P}_{\cal W}(\epsilon)$ otherwise $V=C_{\cal W}({\cal P}_0,{\cal A}_0)$ can be increased by performing another iteration. Specifically, if $V$ is reached within finite iterations $m'$, ${\cal P}_0$ and ${\cal A}_0$  are pairwise optimal for ${\sf P}_{\cal W}(\epsilon_0/i')$; otherwise, ${\cal P}_0$ and ${\cal A}_0$  are pairwise optimal for ${\sf P}_{\cal W}(0)$.

We first consider the case that $V$ is reached within finite iterations $i'$. Since $V$ is reached within finite iterations $i'$, we have that ${\cal P}_0$ and ${\cal A}_0$ are pairwise optimal for both ${\sf P}_{\cal W}(\epsilon_0 /i')$ and ${\sf P}_{\cal W}(\epsilon_0 /(i'-1))$. Then, by Theorem \ref{thm:jointopt}, $({\cal P}_0,{\cal A}_0)$ is the optimal solution to ${\sf P}_{\cal W}(\epsilon_0 /i')$ and ${\sf P}_{\cal W}(\epsilon_0 /(i'-1))$. Note that, since the feasible domain of ${\sf P}_{\cal W}(\epsilon_0 /i')$ is expanded from that of ${\sf P}_{\cal W}(\epsilon_0 /(i'-1))$ by decreasing $\epsilon$, $({\cal P}_0,{\cal A}_0)$ is not on the boundary of $a_n^k \geq \epsilon_0/i'$, i.e., the equality of $a_m^k \geq \epsilon_0/i'$ does not hold. Therefore, continually expanding the feasible domain of ${\sf P}_{\cal W}(\epsilon)$ by decreasing $\epsilon$ from $\epsilon_0/i'$ to $0$, $({\cal P}_0,{\cal A}_0)$ remains at a local optimal point and thus also a global optimal point according to the domain's convexity.

We then consider the case that $V$ can only be approached with infinite iterations. For this case, we have that ${\cal P}_0$ and ${\cal A}_0$ are pairwise optimal for ${\sf P}_{\cal W}(0)$. However, we note that, even so, $({\cal P}_0,{\cal A}_0)$ is not necessarily optimal solution to ${\sf P}_{\cal W}(0)$ when $a_m^k=0$ for some $m\in{\cal M}$ and $k\in{\cal K}$. To show the optimality of $({\cal P}_0,{\cal A}_0)$, we use the proof by contradiction. Suppose that ${\cal P}_0$ and ${\cal A}_0$ are pairwise optimal for ${\sf P}_{\cal W}(0)$ but  $({\cal P}_0,{\cal A}_0)$ is not an optimal solution to ${\sf P}_{\cal W}(0)$. Denote ${\cal Z}\triangleq \{(m,k)\;|\;a_m^k=0, a_m^k\in{\cal A}_0,p_m^k\in{\cal P}_0\}$ as the set of the links with zero bandwidth allocation. Since ${\cal P}_0$ and ${\cal A}_0$ are pairwise optimal for ${\sf P}_{\cal W}(0)$, we have that all K.K.T. conditions hold except for the links $(n,k)\in{\cal Z}$, i.e., excluding the links in ${\cal Z}$, $({\cal P}_0,{\cal A}_0)$ is optimal for ${\sf P}_{\cal W}(0)$ by Theorem \ref{thm:jointopt} (excluding the links in $\cal Z$, the problem ${\sf P}_{\cal W}(0)$ is equivalent to ${\sf P}_{\cal W}(\epsilon')$ where $\epsilon'$ is the remaining smallest bandwidth allocation). However, since we also have that $({\cal P}_0,{\cal A}_0)$ is not optimal for ${\sf P}_{\cal W}(0)$, then we know that $\{a_m^k=0,p_m^k=0\;|\;(m,k)\in{\cal Z}\}$ is suboptimal, i.e., we can always reassign an arbitrary small bandwidth from some non-zero bandwidth link to a zero bandwidth link and then perform ${\sf EP}_n$ to achieve a new objective value which is higher than $V$. Obviously, due to the increase of the objective value, the energy allocation of the link with the newly assigned bandwidth must increase from zero to a positive value after solving ${\sf EP}_n$ with the new bandwidth allocation. Specifically, for a link $(m,k)\in{\cal Z}$, if reassigning an arbitrary small bandwidth can result in the corresponding $p_m^k$ increased form zero to a positive value, we must have $H_m^k > v_n^k - u_n^k$ such that $m\in{\cal M}_n$ by Theorem \ref{thm:wf}, i.e., according to the water-filling solution, $v_n^k-u_n^k$ increases after solving ${\sf EP}_n$ with the new $a_m^k >0$ while the new $p_m^k$ determined by  \eqref{eq:wf} must be positive.

However, in each specific iteration, we have $a_m^k > 0$ and the optimal solution to ${\sf EP}_n$ satisfies $H_m^k \leq v_n^k - u_n^k$ such that $m\in{\cal M}_n$ when $p_m^k = 0$, by \eqref{eq:wf}. Note that, the objective function is continuous and the problem is a convex optimization problem. Then, following the algorithm, when $a_m^k$ converges to zero, we also have $H_m^k \leq v_n^k - u_n^k$ when $p_m^k = 0$, which is contradiction to the above suboptimal assumption. Therefore,  the converged objective value must be the optimal value for problem ${\sf P}_{\cal W}(0)$.




\section{Proof of Proposition \ref{pp:al}}

Note that, initially, $\bar{\cal T}^k$ contains the elements such that $p_n^k=0$ and, obviously, \eqref{eq:bal2} is satisfied. Following the procedure of Algorithm 2, at the end of each iteration, new elements are added to $\bar{\cal T}^k$. Therefore, we need to show that, for any $n\in\bar{\cal T}^k$, (a) \eqref{eq:bal2} is satisfied for $n$ in the iteration when $n$ is added to $\bar{\cal T}^k$; (b) \eqref{eq:bal2} is still satisfied for $n$ in the next  iterations.

We first show that, at the end of each iteration, $n_0\in{\cal V}$, which is newly added to $\bar{\cal T}^k$, satisfies \eqref{eq:bal2} after this move. At the beginning of the iteration, we have the sets  ${\cal T}^k$ and $\bar{\cal T}^k$. Following Algorithm 2, we recalculate $a_n^k$ by \eqref{eq:bal1} for all $n\in {\cal T}^k$ and all $k\in{\cal K}$. After this recalculation, if ${\cal V}$ is non-empty, i.e., there exists $n_0\in{\cal V}$ such that $a_{n_0}^k \leq \epsilon$, we will move $n_0$ from ${\cal T}^k$ to $\bar{\cal T}^k$ at the end of the iteration. Also, we have
\begin{align}
\epsilon\geq a_{n_0}^k &= p_{n_0}^kH_{n_0}^k\frac{1-|\bar{{\cal T}}^k|\cdot\epsilon}{\sum_{i\in{\cal T}^k}p_i^kH_i^k}\label{eq:val0}\\
&=p_{n_0}^kH_{n_0}^k\frac{1-|\bar{{\cal T}}^k|\cdot\epsilon-a_{n_0}^k}{\sum_{i\in{\cal T}^k}p_i^kH_i^k-p_{n_0}^kH_{n_0}^k}\label{eq:val1}\\
&\geq p_{n_0}^kH_{n_0}^k\frac{1-|\bar{{\cal T}}^k|\cdot\epsilon-\epsilon}{\sum_{i\in{\cal T}^k}p_i^kH_i^k-p_{n_0}^kH_{n_0}^k}\label{eq:val2}\\
&= p_{n_0}^kH_{n_0}^k\frac{1-|\bar{{\cal T}}^k\cup n_0|\cdot\epsilon}{\sum_{i\in{\cal T}^k\slash n_0}p_i^kH_i^k},\label{eq:val3}
\end{align}
where \eqref{eq:val1} follows since $a_{n_0}^k = p_{n_0}^kH_{n_0}^k\frac{1-|\bar{{\cal T}}^k|\cdot\epsilon}{\sum_{i\in{\cal T}^k}p_i^kH_i^k}$ and $a_{n_0}^k = p_{n_0}^kH_{n_0}^k \frac{a_{n_0}^k}{p_{n_0}^kH_{n_0}^k}$ and  we know that if $c = (a + b) / (x + y)$ and $c = b / y$, then $c = a / x = b / y = (a + b) / (x + y)$. \eqref{eq:val2} follows since $a_{n_0}^k\leq \epsilon$.

Rearranging \eqref{eq:val3}, we have $\epsilon\geq (1-|\bar{{\cal T}}^k\cup n_0|\cdot\epsilon) \frac{p_{n_0}^kH_{n_0}^k}{\sum_{i\in{\cal T}^k\slash n_0}p_i^kH_i^k}$ where $n_0\in\bar{{\cal T}}^k\cup n_0$, $\bar{{\cal T}}^k\leftarrow \bar{{\cal T}}^k\cup n_0$ and ${\cal T}^k\leftarrow{\cal T}^k\slash n_0$ are the new sets generated at the end of the iteration, respectively. Hence, $n_0$, which is newly added to $\bar{\cal T}^k$, satisfies \eqref{eq:bal2}.

We next show that, $n_0$ will also satisfy \eqref{eq:bal2} in subsequent iterations. By \eqref{eq:val0}-\eqref{eq:val3}, we also have
\begin{equation}
\frac{1-|\bar{{\cal T}}^k|\cdot\epsilon}{\sum_{i\in{\cal T}}p_i^kH_i^k}\geq \frac{1-|\bar{{\cal T}}^k\cup n_0|\cdot\epsilon}{\sum_{i\in{\cal T}^k\slash n_0}p_i^kH_i^k}\ ,
\end{equation}
i.e., the value of $\frac{1-|\bar{{\cal T}}^k|\cdot\epsilon}{\sum_{i\in{\cal T}}p_i^kH_i^k}$ decreases over the iterations, and so is the value of $p_{n0}^kH_{n0}^k\frac{1-|\bar{{\cal T}}^k|\cdot\epsilon}{\sum_{i\in{\cal T}}p_i^kH_i^k}$. Moreover, by \eqref{eq:val0}, we have that $\epsilon\geq p_{n_0}^kH_{n_0}^k\frac{1-|\bar{{\cal T}}^k|\cdot\epsilon}{\sum_{i\in{\cal T}}p_i^kH_i^k}$ in the current iteration. Therefore, in the subsequent iterations, \eqref{eq:bal2} remains satisfied for $n_0$.


%

\ifCLASSOPTIONcaptionsoff
  \newpage
\fi



%

\bibliographystyle{IEEETran}
\bibliography{IEEEabrv,bib}

\begin{thebibliography}{10}
\providecommand{\url}[1]{#1}
\csname url@samestyle\endcsname
\providecommand{\newblock}{\relax}
\providecommand{\bibinfo}[2]{#2}
\providecommand{\BIBentrySTDinterwordspacing}{\spaceskip=0pt\relax}
\providecommand{\BIBentryALTinterwordstretchfactor}{4}
\providecommand{\BIBentryALTinterwordspacing}{\spaceskip=\fontdimen2\font plus
\BIBentryALTinterwordstretchfactor\fontdimen3\font minus
  \fontdimen4\font\relax}
\providecommand{\BIBforeignlanguage}[2]{{%
\expandafter\ifx\csname l@#1\endcsname\relax
\typeout{** WARNING: IEEEtran.bst: No hyphenation pattern has been}%
\typeout{** loaded for the language `#1'. Using the pattern for}%
\typeout{** the default language instead.}%
\else
\language=\csname l@#1\endcsname
\fi
#2}}
\providecommand{\BIBdecl}{\relax}
\BIBdecl

\bibitem{EHSNSI}
S.~Sudevalayam and P.~Kulkarni, ``Energy harvesting sensor nodes: survey and
  implications,'' \emph{{IEEE} Commun. Surveys Tuts.}, vol.~13, no.~3, pp.
  443--461, Sep. 2011.

\bibitem{Energy_Scav}
J.~A. Paradiso and T.~Starner, ``Energy scavenging for mobile and wireless
  electronics,'' \emph{IEEE Trans. Pervasive Computing}, vol.~4, pp. 18--27,
  Jan. 2005.

\bibitem{WPDFUE}
G.~Maria, W.~Aya, and Z.~Gil, ``Networking low-power energy harvesting devices:
  measurements and algorithms,'' \emph{{IEEE} Trans. Mobile Comput.}, vol.~12,
  no.~9, pp. 1853--1233, Sep. 2013.

\bibitem{PMEHW}
J.~Piorno, C.~Bergonzini, K.~Atienza, and T.~Rosing, ``Prediction and
  management in energy harvested wireless sensor nodes,'' in \emph{VITAE 2009},
  May 2009, pp. 6--10.

\bibitem{ALLERTON}
Z.~Wang, V.~Aggarwal, and X.~Wang, ``Renewable energy scheduling for fading
  channels with maximum power constraint,'' in \emph{Allerton 2013}, Oct. 2013,
  pp. 1394--1400.

\bibitem{2014arXiv1401.2376W}
------, ``{Iterative Dynamic Water-filling for Fading Multiple-Access Channels
  with Energy Harvesting},'' \emph{available at arXiv 1401.2376}, 2013.

\bibitem{OTPBLE}
K.~Tutuncuoglu and A.~Yener, ``Optimum transmission policies for battery
  limited energy harvesting nodes,'' \emph{{IEEE} Trans. Commun.}, vol.~11,
  no.~3, pp. 1180--1189, Mar. 2012.

\bibitem{TMGRCE}
C.~Huang, R.~Zhang, and S.~Cui, ``Throughput maximization for the gaussian
  relay channel with energy harvesting constraints,'' \emph{{IEEE} J. Sel.
  Areas Commun.}, vol.~PP, no.~99, pp. 1--11, Sep. 2012.

\bibitem{OPSMAC}
J.~Yang and S.~Ulukus, ``Optimal packet scheduling in a multiple access channel
  with energy harvesting transmitters,'' \emph{{IEEE} J. Commun. and Netw.},
  vol.~14, no.~2, pp. 140--150, Apr. 2012.

\bibitem{FHEARS}
S.~Chen, P.~Sinha, N.~Shroff, and C.~Joo, ``Finite-horizon energy allocation
  and routing scheme in rechargeable sensor networks,'' in \emph{Proc. IEEE
  2011 INFOCOM}, Apr. 2011, pp. 2273--2281.

\bibitem{OEAWCE}
C.~Ho and R.~Zhang, ``Optimal energy allocation for wireless communications
  with energy harvesting constraints,'' \emph{{IEEE} Trans. Signal Process.},
  vol.~60, no.~9, pp. 4808--4818, Sep. 2012.

\bibitem{TEHNFW}
O.~Ozel, K.~Tutuncuoglu, J.~Yang, S.~Ulukus, and A.~Yener, ``Transmission with
  energy harvesting nodes in fading wireless channels: optimal policies,''
  \emph{{IEEE} J. Sel. Areas Commun.}, vol.~29, no.~8, pp. 1732--1743, Sep.
  2011.

\bibitem{SROPPE}
K.~Tutuncuoglu and A.~Yener, ``Sum-rate optimal power policies for energy
  harvesting transmitters in an interference channel,'' \emph{{IEEE} J.
  Commun., Netw.}, vol.~14, no.~2, pp. 151--161, Apr. 2012.

\bibitem{broadcasting}
O.~Ozel, Y.~Jing, and S.~Ulukus, ``Optimal broadcast scheduling for an energy
  harvesting rechargeable transmitter with a finite capacity battery,''
  \emph{{IEEE} Trans. Wireless Commun.}, vol.~11, no.~8, pp. 2193--2203, Jun.
  2012.

\bibitem{CO}
S.~Boyd and L.~Vandenberghe, \emph{Convex Optimization}.\hskip 1em plus 0.5em
  minus 0.4em\relax Cambridge: Cambridge University Press, 2009.

\bibitem{IT}
T.~Cover and J.~Thomas, \emph{Elements of Information Theory}.\hskip 1em plus
  0.5em minus 0.4em\relax New York: Wiley, 1991.

\bibitem{FCPMRA}
A.~Duel-Hallen, ``Fading channel prediction for mobile radio adaptive
  transmission systems,'' \emph{Proc. {IEEE}}, vol.~95, no.~12, pp. 2299--2313,
  Dec. 2007.

\end{thebibliography}

%

%







\end{document}